\documentclass[%
aip,jcp,amsmath,amssymb, reprint, twocolumn
]{revtex4-2}
\usepackage{amsmath}
\usepackage{mathtools}
\usepackage{float}
\usepackage{amsfonts}
\usepackage{amssymb}
\usepackage{dcolumn} %% tables cols aligned at decimal point
\usepackage{array}
\newcolumntype{P}[1]{>{\centering\arraybackslash}p{#1}}
\newcolumntype{M}[1]{>{\centering\arraybackslash}m{#1}}
\newcolumntype{C}[1]{>{\centering\arraybackslwash}p{#1}}
\usepackage{float}
\usepackage{psfrag}
\usepackage{tabularx}
\usepackage{stackengine}
\usepackage{amssymb}
\usepackage{mathtools}  
\usepackage{xfrac} 
\usepackage[T1]{fontenc}
\usepackage{graphicx}
\usepackage{diagbox}
\usepackage[caption=false]{subfig}
\usepackage{tikz}
\usetikzlibrary{positioning}
\usetikzlibrary{arrows}
\usetikzlibrary{trees}
\usepackage{amssymb}
\usetikzlibrary{decorations.pathmorphing}
\usetikzlibrary{decorations.markings}
\usetikzlibrary{automata,positioning}
\usepackage{braket}
\usepackage{rotating}
\usepackage{adjustbox}
\usepackage{ragged2e}
\usepackage{simplewick}
\usepackage{simpler-wick}

\usepackage[]{lineno}

\usepackage{csquotes}
\usepackage{physics,amsmath}
\usepackage{xcolor}
\usepackage{fancyhdr}
\UseRawInputEncoding
\usepackage[normalem]{ulem}

\usepackage{scalerel}
\usepackage{hyperref}
\usepackage{appendix}

\begin{document}

\author{Chayan Patra}
\affiliation{ Department of Chemistry,  \\ Indian Institute of Technology Bombay, \\ Powai, Mumbai 400076, India}

\author{Sonaldeep Halder}
\affiliation{ Department of Chemistry,  \\ Indian Institute of Technology Bombay, \\ Powai, Mumbai 400076, India}

\author{Rahul Maitra}
\email{rmaitra@chem.iitb.ac.in}
\affiliation{ Department of Chemistry,  \\ Indian Institute of Technology Bombay, \\ Powai, Mumbai 400076, India}
\affiliation{Centre of Excellence in Quantum Information, Computing, Science \& Technology, \\ Indian Institute of Technology Bombay, \\ Powai, Mumbai 400076, India}
%\title{A Resource Efficient Projective Quantum Eigensolver via Adiabatically Decoupled Subsystem Evolution}
\title{Projective Quantum Eigensolver via Adiabatically Decoupled Subsystem Evolution: a Resource Efficient Approach to Molecular Energetics in Noisy Quantum Computers}
%\linenumbersCorrections

\begin{abstract}
Quantum computers hold immense potential in the field of chemistry, ushering new frontiers to solve complex many body problems that are beyond the reach of classical computers. However, noise in the current quantum hardware limits their applicability to large chemical systems. This work encompasses the development of a projective formalism that aims to compute ground-state energies of molecular systems accurately using Noisy Intermediate Scale Quantum (NISQ) hardware in a resource efficient manner. Our approach is reliant upon the formulation of a bipartitely decoupled
parameterized ansatz within the disentangled unitary coupled cluster (dUCC) framework based on the principles of \textit{synergetics}.
Such decoupling emulates the total parameter optimization in a lower dimensional manifold, while a mutual synergistic relationship
among the parameters is exploited to ensure characteristic accuracy. Without any \textit{pre-circuit measurements}, our method leads to a highly compact fixed-depth ansatz with
shallower circuits and fewer expectation value evaluations. Through analytical and numerical demonstrations, we demonstrate the method's superior performance under noise while concurrently ensuring requisite accuracy in future fault-tolerant systems. This approach enables rapid exploration of emerging chemical spaces by efficient utilization of near-term quantum hardware resources.

\end{abstract}

\maketitle

\section{Introduction}
 The exact description of many body systems evolve in an
exponentially growing Hilbert space (with system size), leading to
eventually an intractable problem\cite{kohn1999nobel} and restricting the conventional
computers to rely heavily on mean-field approaches for real-life
applications. However, these mean-field approaches often inherently ignore
the important physics, such as the correlation effects within an ensemble
of interacting particles. For atomic and molecular systems, these electron-electron
correlation effects are of paramount importance in
chemistry and condensed matter physics and are taken care of by theories like
density matrix renormalization group (DMRG)\cite{white1992density},
selected configuration interaction
\cite{huron1973iterative,buenker1974individualized} and different variants
of the coupled cluster (CC)
methods\cite{cc3,cc4,cc5,bartlett2007coupled,crawford2000introduction}.
Despite being able to predict the energies accurately, these methods have an exponential or higher-order polynomial
cost, hindering their usage for most of the practical applications. 

Quantum computers hold great promise in the field of quantum chemistry, particularly for simulating the electronic behavior of molecules
and materials in an exponentially more efficient manner. Such platforms leverage principles of superposition and 
entanglement to store and manipulate many-fermion wavefunctions.  Early quantum algorithms, including
those employing Quantum Phase Estimation (QPE) \cite{abrams1997simulation,abrams1999quantum}, played a 
pioneering role in this domain. However, these methods rely on deep 
quantum circuits, which may surpass the coherence times available in 
current quantum hardware\cite{lidar1998decoherence, divincenzo2000physical, ladd2010quantum}. Additionally, these quantum devices suffer from poor gate fidelity, state preparation and measurement errors (SPAM), and readout errors. As a result, computations performed using deep quantum circuits are prone to significant inaccuracies. To overcome these challenges, methods 
have been developed that require comparatively shallower circuits and have inherent noise resilience. 
Variational Quantum Eigensolver \cite{peruzzo2014variational} (VQE) is one such 
method that conforms to the limitations of the 
current noisy quantum hardware. It prepares a 
parameterized state within the quantum computing 
framework using a shallow depth circuit (ansatz). The optimum parameters are obtained through 
classical variational minimization of the energy expectation value. Several other variants \cite{grimsley2019adaptive,delgado2021variational,mondal2023development,feniou2023overlap,zhao2023orbital, tang2021qubit, yordanov2021qubit, ostaszewski2021structure, tkachenko2021correlation, zhang2021adaptive, sim2021adaptive,sonaldeep_rbm,halder2022dual,halder2023corrections,halder2024noise}
of VQE framework have been proposed to further minimize the ansatz depth.
Another promising method, the Projective Quantum 
Eigensolver  (PQE)\cite{stair2021simulating, halder2023development, misiewicz2023implementation} also relies on generating a
parameterized quantum state where the parameters are 
optimized by solving a set of nonlinear projective equations. This method
is also shown to exhibit inherent noise resilience and has a faster 
convergence (specifically under noise) as compared to VQE 
\cite{stair2021simulating}. Moreover, it does not show the barren plateau
problem \cite{cerezo2021cost, wang2021noise},  
often encountered during variational energy minimization in the VQE 
framework. PQE can be used with disentangled unitary coupled cluster 
ansatz (dUCC)\cite{evangelista2019exact}, which is capable of 
capturing a substantial amount of correlation energy of molecular
systems. When coupled with a CNOT
efficient implementation\cite{yordanov2020efficient,magoulas2023cnot}, dUCC offers a flexible and efficient ansatz
structure for application in molecular systems.

% However,
% implementation of PQE for large molecules employing the 
% dUCC ansatz demands the utilization of significantly deep quantum circuits. While these circuits may fall within the coherence time, the 
% inherent noise in the current hardware can potentially introduce 
% substantial errors in the final computed ground state energies. 
% Furthermore, the PQE approach mandates a considerable number of 
% measurements throughout its iterative process, which could lead to 
% prolonged runtimes.
Nevertheless, when dealing with large and highly correlated molecules, implementing dUCC demands deep quantum circuits and a large number of optimizable parameters. While increased depth leads to large errors in computed energetics for practical applications, more parameters require many quantum measurements during the PQE iterative method. Hence, there is a pressing need to develop methods capable of overcoming these challenges to attain accurate molecular energetics within the limited coherence time of noisy quantum hardware.
% In this work, we leverage the synergy present within the non-linear PQE iterative process to derive a method that significantly reduces gate depth and minimizes the need for quantum measurements.

In this work, we have developed a \enquote{two-phase} formalism in the PQE framework with the aim of reducing the quantum resources (\textit{both} circuit
gate depth and measurements) utilized during the non-linear iterative optimization. The approach is based on the judicious partitioning of the PQE
parameter space into a smaller dimensional \textit{principal} parameter subspace and a larger dimensional \textit{auxiliary}
subspace.
% These two subspaces have distinctly different timescales of equilibration, allowing us to adiabatically decouple the 
% fast relaxing \textit{auxiliary} modes from the slow relaxing \textit{principal} modes \cite{haken1982slaving,haken2013synergetics}.
% The authors had previously substantiated the existence of such \textit{principal} subspace which macroscopically emulates the
% optimization trajectory while the complementary \textit{auxiliary subspace} gets synergistically 
% coupled via a feedback loop to the former
% \cite{agarawal2020stability,agarawal2021accelerating,agarawal2021approximate,agarawal2022hybrid,patra2023synergistic,halder2023machine}.
The authors had previously substantiated that these two subspaces have distinctly different timescales of equilibration, allowing us to adiabatically
decouple\cite{haken1975generalized,haken2013synergetics,wunderlin1981generalized} the fast relaxing \textit{auxiliary} modes from the slow relaxing \textit{principal} modes\cite{agarawal2020stability,agarawal2021accelerating,agarawal2021approximate,agarawal2022hybrid,patra2023synergistic,halder2023machine,mehendale2023exploring,pathirage2024exploration}.
Consequently, \textit{auxiliary} parameters can be expressed as a mathematical function of the \textit{principal} parameters that projects
the parameter optimization in a lower dimensional manifold.
This is popularly known in literature as the \textit{slaving principle}\cite{haken1982slaving,haken1985application,haken1983nonlinear,haken1983self,wu2000suppressing,iskra1993self,zheng2024order}.
% the \textit{independent variation} of the \textit{auxiliary} parameters as the system of nonlinearly coupled parameters evolves\cite{agarawal2020stability,agarawal2021accelerating,agarawal2021approximate,agarawal2022hybrid,patra2023synergistic,halder2023machine}.
% such a synergistic relationship between separate sets of parameters allows us to neglect the \textit{independent variation} of the \textit{auxiliary} parameters as the system of nonlinearly coupled parameters evolves\cite{agarawal2020stability,agarawal2021accelerating,agarawal2021approximate,agarawal2022hybrid,patra2023synergistic,halder2023machine}.
% Such variation 
% of the auxiliary subspace is distinctively suppressed compared to that exhibited by the principal subspace.
We use this decoupling paradigm within the PQE non-linear optimization scheme and analytically establish the function that maps elements of
\textit{auxiliary} space from the \textit{principal} one. Taking a step further, we introduce 
a \textit{no-feedback-control} formalism, where we capture the effective dependence of the
\textit{auxiliary} parameters on the \textit{principal} ones through a one-step \textit{post-optimization mapping}. 
This allows us to perform PQE
within the lower dimensional and fully disjoint \textit{principal} subspace.
% without even bothering about the 
% \textit{feedback dynamics}\cite{haken1982slaving,agarawal2021approximate,patra2023synergistic,halder2023machine} of the \textit{auxiliary} modes. 
Previous studies\cite{halder2023machine} focusing on subspace decoupling within the PQE formalism showed promise but depended on machine learning and assumed an ideal, noise-free quantum device. Additionally, the approach only resulted in measurement reduction. Other dimensionality reduction techniques such as the subsystem embedding subalgebra coupled cluster (SES-CC)\cite{kowalski2018properties,bauman2019downfolding} and their quantum
computing counterpart\cite{kowalski2021dimensionality,bauman2022coupled,kowalski2023quantum} have also been previously explored, but they operate on different underlying principles. This manuscript delineates the first attempt to formulate and demonstrate the effectiveness of an analytic adiabatic subspace decoupling to come up with an optimally resource-efficient quantum algorithm toward realistic implementation under hardware noise.

Our approach serves twin desirable goals -- while allowing
us to perform PQE with a fixed structured shallow circuit, it also markedly
reduces measurement costs during optimization. This becomes even more advantageous on noisy quantum devices as the 
reduction in the circuit depth entails less noise and shows better effectiveness towards 
various error mitigation strategies. Moreover, the compactification of the ansatz does not involve any \textit{pre-circuit measurement} (measurements prior to the construction of the final ansatz circuit). Thus, the structural optimality of the ansatz is not deteriorated by noise. Molecular energy calculations on NISQ hardware are typically noise-dominated rather than correlation-dominated.
% Therefore, shallower circuits, often insufficiently parameterized, tend to yield more accurate results. However, this apparent accuracy improvement with shallower circuits on NISQ devices can be misleading.
% A significantly shallow ansatz, while appearing accurate on NISQ devices, may not adequately capture the underlying electronic correlation. This limitation becomes apparent as noise levels decrease.
As a consequence, significantly shallower circuits, though seemingly accurate on NISQ devices, might not fully capture electronic correlation due to insufficient parameterization. This limitation is likely to become more apparent as the noise in quantum hardware is reduced. The hallmark of a robust ansatz is 
thus its ability to balance expressibility and agility under noise. In this regard, an open and nontrivial question
that we numerically address in this manuscript is the choice of the
optimal dimension for the \textit{principal} subspace that balances
the two opposing effects, such that the theory which is 
reasonably accurate in the NISQ era can also be seamlessly adopted
when better hardware becomes available. This flexibility enables 
our method to be used under various hardware infrastructures with varying 
degrees of noise, facilitating the exploration of upcoming emerging chemical 
phenomena in a hardware-agnostic way.

The manuscript is structured as follows: we begin by formalizing the \textit{no-feedback-controlled} adiabatically decoupled PQE (\textit{nfc}AD-PQE). Following this, 
we present a mathematical foundation establishing its resource efficiency and utility in noisy quantum hardware.
% , showcasing enhanced error mitigability.
Subsequently, we demonstrate its efficiency and accuracy in an ideal, noiseless
scenario. In the final section, we illustrate the efficacy of \textit{nfc}AD-PQE under noise and compare it with conventional PQE.

\section{Theoretical Development of no-feedback-controlled Adiabatically Decoupled Projective Quantum Eigensolver}
\subsection{Emergence of Collective Behaviour: Slaving Principle in Projection Based Unitary Coupled Cluster Theory} \label{general math prelim}

% \begin{figure}[!h]
%     \centering
% \includegraphics[width=\linewidth]{sorted_params.png}
% \caption{}
%     \label{iter_conv_plot}
% \end{figure}

% Unitary coupled cluster(UCC) is an alternate CC formalism with the ansatz-

Projective Quantum Eigensolver (PQE)\cite{stair2021simulating} relies on the preparation of a parameterized trial wavefunction $\ket{\Psi(\theta)}$ by the action of a unitary $\hat{U}(\theta)$ on a reference state $\ket{\Phi_o}$, often taken to be the Hartree Fock (HF) state

\begin{equation} \label{psi}
   \ket{\Psi (\boldsymbol{\theta})} = \hat{U}(\boldsymbol{\theta})\ket{\Phi_o}
\end{equation}
Here, in the disentangled Unitary Coupled Cluster (dUCC) ansatz \cite{evangelista2019exact}, $\hat{U}$ is taken to be - 

\begin{equation}\label{unitary lexical}
    \hat{U}(\boldsymbol{\theta}) = \prod_{\mu} e^{\theta_{\mu}\hat{\kappa}_{\mu}}
\end{equation}
where, $\hat{\kappa}_{\mu}=\hat{\tau}_{\mu}-\hat{\tau}_{\mu}^\dagger$ is an anti-Hermitian operator with $\hat{\tau}_{\mu}= \hat{a}^{\dagger}_{a}\hat{a}^{\dagger}_{b}....\hat{a}_{j}\hat{a}_{i}$ and the boldface $\boldsymbol{\theta}$ represents the set
of all parameters in a vectorized form. 
% \begin{equation} \label{kappa_mu_def}
%     \hat{\kappa}_{\mu}=\hat{\tau}_{\mu}-\hat{\tau}_{\mu}^\dagger
% \end{equation}    
% \begin{equation}\label{creat_annhil_def}
%     \hat{\tau}_{\mu}= \hat{a}^{\dagger}_{a}\hat{a}^{\dagger}_{b}....\hat{a}_{j}\hat{a}_{i}
% \end{equation}
In the above equations, $\mu$ represents a multi-index particle-hole excitation structure as defined by the string of creation ($\hat{a}^{\dagger}$) and annihilation ($\hat{a}$) operators with the indices $\{i,j,...\}$ denoting the occupied spin-orbitals in the HF state and $\{a,b,...\}$ denoting the unoccupied spin-orbitals.

% $\hat{U}(\boldsymbol{\theta}) = \prod_{\mu}
% e^{\theta_{\mu}\hat{\kappa}_{\mu}}$. It is an (exponentiated) ordered product of the non-commuting
% $\hat{\kappa}_{\mu}$ operators for a composite hole-particle index $\mu$, $\theta_{\mu}$ are the
% parameters corresponding to the anti-hermitian operator $\hat{\kappa}_{\mu}=\hat{T}_{\mu}-\hat{T}_{\mu}^\dagger$. 
% % UCC is not feasible for classical algorithm due to the non-termination
% of BCH expansion of similarity transformed Hamiltonian while such unitary gate operations are natural
% in quantum computing framework. 

% The recently proposed projective quantum eigensolver (PQE)\cite{stair2021simulating} considers this as a

PQE determines the optimal parameters by considering it as a nonlinear optimization problem, which is accomplished iteratively via residue construction- 

\begin{equation} \label{iter}
    \theta_{\mu}^{(k+1)} = \theta_{\mu}^{(k)} + \frac{r_{\mu}^{(k)}}{D_{\mu}}.
\end{equation}
where, \textit{k} represents an iterative step with $D_{\mu}$ denoting the second-order Moller Plesset (MP2) denominator\cite{moller1934note}. The residue ($r_{\mu}$) is obtained by projecting the Schrodinger equation against excited determinants $\{\ket{\Phi_{\mu}}\}$ -
\begin{equation} \label{4}
       r_{\mu}(\boldsymbol{\theta}) = \bra{\Phi_{\mu}}\hat{U}^{\dagger}(\boldsymbol{\theta})H \hat{U}(\boldsymbol{\theta})\ket{\Phi_o} ; \mu \ne 0
\end{equation}
$r_{\mu}$ can be executed on a quantum computer as a sum of three expectation value calculations\cite{stair2021simulating}-
\begin{equation}
    r_{\mu} = \bra{\Omega_{\mu}(\sfrac{\pi}{4})} \bar{H} \ket{\Omega_{\mu}(\sfrac{\pi}{4})} - \frac{1}{2}E_{\mu} - \frac{1}{2}E_0
\end{equation}
with the following definitions, $\ket{\Omega_{\mu}(\theta)} = e^{\hat{\kappa}_{\mu} \theta} \ket{\Phi_0}$, $\bar{H} = \hat{U}^{\dagger}\hat{H}\hat{U}$, $E_\mu = \bra{\Phi_\mu} \bar{H} \ket{\Phi_\mu}$
and $E_0 = \bra{\Phi_0} \bar{H} \ket{\Phi_0}$.
The iterative procedure is deemed converged when
\begin{equation} \label{convergence condition}
    r_{\mu}(\boldsymbol{\theta^{*}}) \rightarrow 0\mbox{ (at fixed point } \boldsymbol{\theta^{*}})
\end{equation}

However, in examining the discrete-time iteration dynamics (Eq. \eqref{iter}), a discernible hierarchical structure emerges regarding the convergence timescale of the parameters. Notably, parameters of larger magnitudes require a significantly larger number of iterations to achieve convergence compared to their smaller magnitude counterparts\cite{halder2023machine}. In general, for a certain class of multivariable nonlinearly coupled dynamical systems, such co-existing hierarchy among the components can be
exploited to eliminate the fast-converging variables\cite{van1985elimination}. This method projects the system in a lower dimensional manifold in the phase space that indeed simplifies the complex motion to a great extent. For a mathematical analysis, let us cast the parameter update
equation in a generic discrete-time dynamical form by reordering Eq. \eqref{iter} and expanding the residual vector $r_\mu$ into
linear and nonlinear parts-

\begin{equation}
    \Delta{\theta_{\mu}} = \Lambda_\mu \theta_\mu + M(\boldsymbol{\theta})
\end{equation}
where, $\Delta$ is a \textit{difference operator} in a discretized time domain such that $\Delta{\theta_{\mu}} = \theta_{\mu}^{(k+1)} - \theta_{\mu}^{(k)} $, $\Lambda_\mu$ is the coefficient of the diagonal linear term and $M(\boldsymbol{\theta})$
encapsulates the off-diagonal linear terms and all nonlinear couplings among the parameters. For a qualitative phase-space analysis
of such a nonlinear system, one can perturb the system slightly from a fixed point and
perform a \textit{linear stability analysis}\cite{strogatz2018nonlinear,agarawal2020stability} that divides the variables
into two parts based on the following conditions on the eigen
values ($\lambda$) of the \textit{stability matrix}

\begin{equation} \label{eigen values}
\begin{split}
   \{ {\lambda}_P \} <  0 ; & \hspace{5mm} \mbox{corresponds to slow eigendirections: principal} \\ 
  & \hspace{5mm} \mbox{modes (total number }N_P) \\
   \{ {\lambda}_A \} < 0; &      \hspace{5mm} \mbox{corresponds to fast eigendirections: auxiliary} \\ 
  & \hspace{5mm} \mbox{modes (total number } N_A)
\end{split}
\end{equation}
with the very important condition: 
\begin{equation} \label{lambda_condition}
    \mid \lambda_{A}\mid > \mid \lambda_{P} \mid     
\end{equation}
for the fixed point to be a \textit{stable node} \cite{strogatz2018nonlinear}. Even though diagonalizing the \textit{stability matrix} is a
standard procedure in nonlinear dynamics, it is often computationally expensive\cite{szakacs2008stability}, rendering it an impractical approach for the segregation.
This necessitates the decoupling
of the parameter space on the temporal hierarchy of the convergence timescale, which is
computationally less expensive and fairly accurate (if not exact) with the following distinctive
features\cite{haken2004introduction,agarawal2021approximate,patra2023synergistic,halder2023machine}-
\begin{itemize}
    \item Principal Parameter Subset (\textit{PPS}) \{$\theta_P$\}: Usually larger in magnitude, takes much more number of iterations to
    converge and a smaller subset. 
    % \textcolor{red}{The disentangled unitary formed by the set of
    % these principal parameters may be referred to as \textit{Principal Sub-system Unitary}.}
    \item Auxiliary Parameter Subset (\textit{APS}) \{$\theta_A$\}: Smaller in magnitude, converges much faster and a larger subset.
\end{itemize}
such that
\begin{equation} \label{number of var}
    N_P << N_A
\end{equation}
whereas for the magnitude-based decoupling to hold, we require
\begin{equation} \label{magnitude params}
    \lvert \theta_P \rvert >>  \lvert \theta_A \rvert
\end{equation}
Here we define the \textit{principal parameter fraction} $f_{pps}$ such that
\begin{equation}
    f_{pps} = \frac{N_P}{N_{par}}
\end{equation}
where, $N_{par}$ is the total number of parameters. With the notion that such implicit partitioning of the parameter space exists \cite{agarawal2021approximate,patra2023synergistic,halder2023machine},
we may recast the disentangled unitary into two sub-parts $\hat{U}_A$ and $\hat{U}_P$ such that it forms a composite \textit{principal-auxiliary bipartite} unitary ($\hat{U}_{pab}$) operator 
% \begin{equation} \label{mag based U}
%     \begin{split}
%         \hat{U}_{pab} &= \hat{U}_P \hat{U}_A \\
%         & = \Big( (e^{\theta_{P_1}\hat{\kappa}_{P_1}} \cdot e^{\theta_{P_2}\hat{\kappa}_{P_2}} ....) \Big)
%         & = \prod_{K}^{N_P} e^{\theta_{P_K}\hat{\kappa}_{P_K}} \prod_{\beta}^{N_A}  e^{\theta_{A_\beta}\hat{\kappa}_{A_\beta}}
%     \end{split}
% \end{equation}

\begin{equation} \label{mag based U}
\begin{split}
    & \hat{U}_{pab} = \hat{U}_P \cdot \hat{U}_A \\
    & = \prod_{K=1}^{N_P} e^{\theta_{P_K}\hat{\kappa}_{P_K}} \prod_{\beta=1}^{N_A}  e^{\theta_{A_\beta}\hat{\kappa}_{A_\beta}} \\
    & = \Big[ \Big(e^{\theta_{P_1}^D\hat{\kappa}_{P_1}^D} ... e^{\theta_{P_I}^D\hat{\kappa}_{P_I}^D} \cdot e^{\theta_{P_J}^D\hat{\kappa}_{P_J}^D}....\Big) \cdot \Big(e^{\theta_{P_1}^S\hat{\kappa}_{P_1}^S}\\
    & ... e^{\theta_{P_K}^S\hat{\kappa}_{P_K}^S} \cdot e^{\theta_{P_L}^S\hat{\kappa}_{P_L}^S}....\Big)\Big] \Big[ \Big(e^{\theta_{A_1}^D\hat{\kappa}_{A_1}^D} ... e^{\theta_{A_\alpha}^D\hat{\kappa}_{A_\alpha}^D}\\
    & \cdot e^{\theta_{A_\beta}^D\hat{\kappa}_{A_\beta}^D}....\Big) \cdot \Big(e^{\theta_{A_1}^S\hat{\kappa}_{A_1}^S}... e^{\theta_{A_\gamma}^S\hat{\kappa}_{A_\gamma}^S} \cdot e^{\theta_{A_\delta}^S\hat{\kappa}_{A_\delta}^S}....\Big)\Big]
\end{split}
\end{equation}
where, $\theta_*^{(S/D)}$ indicates parameters corresponding to singles or doubles excitations and $P$($A$) are the
\textit{principal} (\textit{auxiliary}) subscripts respectively. In the above expansion, both \textit{principal} ($\hat{U}_P$) and \textit{auxiliary} ($\hat{U}_A$) subsystem unitary operators are ordered according to their
excitation rank (singles or doubles). Within each rank, the individual operators are arranged in regard to the relative absolute magnitude of the corresponding parameters such that
in Eq. \eqref{mag based U} we have $\theta_{P_I}^D > \theta_{P_J}^D$, $\theta_{P_K}^S > \theta_{P_L}^S$ and so on.
The benefit of such an ordering will eventually be pronounced in the numerical analysis and comparative studies.
% \textcolor{red}{Note that $U_{pab}$ in Eq. \eqref{mag based U} forms a magnitude-based ordered 
% set of exponentials (rather than lexicographic ordering as appears in Eq. \eqref{unitary lexical})
% where the principal and the auxiliary operators are separately sub-grouped.
Such parameter space decoupling leads to the discrete-time \textit{equations of motion} of these 
two subsets as follows:
\begin{equation} \label{del_thetaS}
\begin{split}
    \Delta \theta_{A_\alpha} &=  \Lambda_{A_\alpha} \theta_{A_\alpha} + M_{A_\alpha}(\boldsymbol{\theta_P},\boldsymbol{\theta_A}); \forall \alpha= 1, 2, ..., N_A
\end{split}
\end{equation}
and
\begin{equation} \label{del_tL}
\begin{split}
    \Delta \theta_{P_I} &= \Lambda_{P_I} \theta_{P_I}+ M_{P_I}(\boldsymbol{\theta_P},\boldsymbol{\theta_A}); \forall I= 1, 2, ..., N_P
\end{split}
\end{equation}
% The diagonal terms have the following general form:
% \begin{equation} \label{lambda def}
% \begin{split}
%     & \Lambda_{\xi} = \frac{1}{D_{\xi}} \bra{\Phi_{\xi}}  [\hat{H},\hat{\kappa}_{\xi}] \ket{\Phi_{0}}\\
%     %& \Lambda_{P_I} = \frac{1}{D_{P_I}} \bra{\Phi_{P_I}}  [\hat{H},\hat{\kappa}_{P_I}] \ket{\Phi_{0}}
% \end{split}
% \end{equation}
Here, $\Lambda_{\xi}= \frac{1}{D_{\xi}} \bra{\Phi_{\xi}}  [\hat{H},\hat{\kappa}_{\xi}] \ket{\Phi_{0}}$ are the diagonal coefficient terms
and $M_{\xi}(\boldsymbol{\theta_P},\boldsymbol{\theta_A})$ contain all the off-diagonal and nonlinearly
coupled terms, where, depending on the context $\xi \in$ \textit{PPS} or \textit{APS} (see Appendix \ref{validity_ad_approximation} for
explicit expressions).
% \begin{equation} \label{P}
%     \begin{split}
%         &M_{\xi}(\{\theta_{P},\theta_{A}\}) = \frac{1}{D_{\xi}} \bra{\Phi_{\xi}} \hat{H} + \sum_{\nu \neq \xi} \theta_{\nu} [\hat{H},\hat{\kappa}_{\nu}] \\
%  & + \sum_{\nu} \sum_{\mu > \nu} \theta_{\nu}
%  \theta_{\mu} \Big[[\hat{H},\hat{\kappa}_{\nu}],\hat{\kappa}_{\mu}\Big] + . . . \ket{\Phi_0}
%     \end{split}
% \end{equation}
% Note that in both the equations above, we have not yet imposed any restrictions on the labels $\mu,\nu$.
% The most general solution to the \textit{auxiliary} parameter equation Eq. \eqref{del_thetaS} can be immediately written as \cite{haken1997discrete}
% \begin{equation}\label{theta_Ai gen}
%     \theta_{A_\alpha} = \sum_{m=-\infty}^{l} (1+ \Lambda_{A_\alpha})^{l-m} M_{A_\alpha}
% \end{equation}
% By definition $\Lambda_{A_\alpha}$ (Eq. \eqref{lambda def}) is in general negative (see Appendix \ref{app:section1}), ensuring the convergent series expansion of the \textit{auxiliary} modes.
With the conjecture that a hierarchy regarding the timescale of convergence exist among the parameters, the time variation of the \textit{auxiliary} parameters can be neglected
(i.e. $\Delta{\theta_{A_\alpha}}=0$ in Eq. \eqref{del_thetaS})
via \textit{adiabatic approximation}\cite{haken1983nonlinear,agarawal2021approximate,patra2023synergistic,van1985elimination}.
This approximation is valid within the characteristic timescale of
convergence of the \textit{principal} parameters.
Starting from Eq. \eqref{del_thetaS} we can employ \textit{adiabatic approximation} and relative magnitude condition (Eq. \eqref{magnitude params}) to immediately
express the larger dimensional \textit{auxiliary} parameters as a function of \textit{principal} parameters only, which is succinctly captured in
% \begin{equation} \label{ts_gen}
%     \theta_{A_\alpha} = -\frac{M(\{\theta_P,\theta_A\})}{\Lambda_{A_\alpha}} %-\frac{\Delta M(\{\theta_P,\theta_A\})}{\Lambda_{A_\alpha}^2}    
% \end{equation}
% Further, the relative magnitude condition (Eq. \eqref{magnitude params}) can be invoked to express the \textit{auxiliary} parameters as a function of \textit{principal} parameters only, which is succinctly captured in
% \begin{equation} \label{thetaS_ad}
%     \begin{split}
%         \theta_{A_\alpha} &= {-\frac{M_{A_\alpha}(\{\theta_P\})}{\Lambda_{A_\alpha}}}\\
%         & = -\frac{\bra{\Phi_{A_\alpha}} \hat{U}_{P}^{\dagger}(\{\theta_P\}) \hat{H} \hat{U}_{P}(\{\theta_P\}) \ket{\Phi_0}}{\bra{\Phi_{A_\alpha}}  [\hat{H},\hat{\kappa}_{A_\alpha}] \ket{\Phi_{0}}} \\
%         & = \frac{\bra{\Phi_{A_\alpha}} \hat{U}_{P}^{\dagger}(\{\theta_P\}) \hat{H} \hat{U}_{P}(\{\theta_P\}) \ket{\Phi_0}}{D_{A_\alpha}}
%     \end{split}
% \end{equation}
\begin{equation} \label{thetaS_ad}
    \begin{split}
        \theta_{A_\alpha} &= -\frac{M(\boldsymbol{\theta_P})}{\Lambda_{A_\alpha}}= \frac{\bra{\Phi_{A_\alpha}} \hat{U}_{P}^{\dagger}(\boldsymbol{\theta_P}) \hat{H} \hat{U}_{P}(\boldsymbol{\theta_P}) \ket{\Phi_0}}{D_{A_\alpha}}
    \end{split}
\end{equation}
where,
\begin{equation} \label{U_P}
    \hat{U}_{P}(\boldsymbol{\theta_P}) = \prod_{I}^{N_P} e^{\theta_{P_I}\hat{\kappa}_{P_I}}
\end{equation}
is a \textit{principal subsystem unitary} operator. The foundation of projecting the iterative trajectory into a lower-dimensional subspace hinges on the interrelationship depicted in Equation \eqref{thetaS_ad}.
% , and ${\bra{\Phi_{A_\alpha}}  [\hat{H},\hat{\kappa}_{A_\alpha}] \ket{\Phi_{0}}} \approx -D_{A_\alpha}$ (see Appendix \eqref{app:section1} for the validity of the approximation).
The solution for \textit{auxiliary} parameters obtained from utilizing the \textit{adiabatic approximation} provides a leading order approximation to the
most general solution of Eq. \eqref{del_thetaS}. The detailed derivation is shown in Appendix \eqref{validity_ad_approximation}.

Having established the fact that the \textit{auxiliary} amplitudes can be
written solely as the function of the \textit{principal} parameters upon the \textit{adiabatic approximation},
one is essentially left with two choices. The most common practice in this scenario 
would be to invoke a \textit{feedback
coupling} where the mapped \textit{auxiliary} parameters (at each optimization step) 
contribute to the equations of the \textit{principal} residues (or equivalently, \textit{principal} 
parameters) at the next iteration via a \textit{circular causality loop}\cite{agarawal2021accelerating,agarawal2021approximate,patra2023synergistic,halder2023machine}. While this method enables the estimation of \textit{auxiliary} parameters using substantially shallower circuits, it requires the realization of a deep quantum circuit involving both \textit{principal} and \textit{auxiliary} parameters for \textit{principal} residue constructions viz. $r_{P_I} = \bra{\Phi_{P_I}} \hat{{U}}_{pab}^\dagger \hat{H} \hat{{U}}_{pab}
\ket{\Phi_0}$.
An alternative approach in this direction would be to introduce a \textit{no-feedback-controlled}
equation for the \textit{principal} amplitudes, which assumes that the optimization space spanned 
by the \textit{principal} parameters is entirely decoupled from the rest. One, however, still 
needs to develop appropriate energy functional to account for the \textit{auxiliary} parameters
which are recessive to the optimization trajectory but have a significant contribution 
to the correlation energy. In the following section, we will invoke the 
\textit{no-feedback-controlled} \textit{principal} subspace evolution and 
would subsequently develop the theoretical framework that maps the evolved \textit{principal}
parameters towards a single-step updation of the \textit{auxiliary} parameters for quantitative accuracy.

\subsection{No-feedback-controlled Evolution: Towards a shallow depth quantum algorithm}
The above discussion describes how the usual synergistic \textit{feedback-coupling} entails all the parameters
for the \textit{principal} residual, resulting in deeper circuit constructions.
To circumvent this issue, we make a rather non-conservative assumption of 
\textit{no-feedback-controlled evolution} of the \textit{principal} 
parameters where the \textit{auxiliary} parameters are assumed to have \textit{no impact} on the optimization trajectory of the former.
% The validity of this assumption is based on particularly two characteristics of this class of nonlinear systems.
The validity for this assumption rests on two key attributes inherent to this class of nonlinear systems.
Firstly, the \textit{fixed point} $\boldsymbol{\theta^*}$ of the iterative trajectory is a \textit{stable node} and an \textit{attracting fixed point}\cite{strogatz2018nonlinear}. In other words, any initial point
starting within the \textit{basin of attraction}\cite{strogatz2018nonlinear} approach the \textit{fixed point} i.e. $\boldsymbol{\theta_l} \rightarrow \boldsymbol{\theta^*}$ as $l \rightarrow \infty$.
Secondly, the \textit{auxiliary} parameters converge almost immediately and 
linger around the phase space trajectory of the \textit{principal} amplitudes till the residual 
condition is satisfied\cite{van1985elimination}.
These two properties justify considering the \textit{fixed point} in the \textit{principal} parameter phase space as approximately equivalent to the fixed point of the entire parameter space. As our interest solely lies in the fixed point properties
of the iterative dynamics, the \textit{no-feedback-controlled} approximation 
would emulate the overall dynamics quite accurately in the lower dimensional subspace of the \textit{principal} parameters with \textit{auxiliary} parameters
mapped only at the convergence. These characteristics collectively ensure 
that a judiciously chosen \textit{principal}
subspace eventually lands on the \textit{fixed point} in its phase space irrespective of the \textit{feedback mechanism}. It should be emphasized that without the \textit{feedback coupling},
proper choice of the \textit{principal} subspace is extremely important. Otherwise, it might lead to imprecise evolution and inaccurate
\textit{auxiliary} parameter mapping.

Assuming that one identifies the appropriate set of \textit{principal} parameters, 
one may construct the residue vector corresponding
to the
% associated \textit{parameters} within  a
subspace spanned solely by the \textit{principal} parameters, disregarding the contributions from the \textit{auxiliary} parameters:
\begin{equation} \label{no feedback r_PI}
    \begin{split}
        r_{P_I} \approx & \bra{\Phi_{P_I}} \hat{U}_{P}^\dagger \hat{H} \hat{U}_{P} \ket{\Phi_0}
    \end{split}
\end{equation}
At this point, it is worth noting the difference between the \textit{no-feedback-controlled} generation of the \textit{principal} residue vectors and the \textit{feedback-controlled} generation: in the case of the latter, one needs to construct the effective Hamiltonian where both \textit{principal} and \textit{auxiliary} parameters contribute, resulting in a proliferation of the 
circuit depth. This is prudently bypassed in the \textit{no-feedback-controlled} generation as exemplified in Eq. \eqref{no feedback r_PI}.

With the optimized \textit{principal} parameters, one may determine the \textit{auxiliary} 
parameters through Eq. \eqref{thetaS_ad} post optimization. One begging question at this
juncture is to develop an appropriate energy expression that can in principle account 
for the correlation energy accurately at the cost of a shallow \textit{fixed-depth-circuit}.
With the assumption that the \textit{principal} and \textit{auxiliary} subspaces remain entirely uncoupled, 
the contributions of the \textit{auxiliary} parameters to energy are to be treated as additional
extraneous corrections. Leveraging the \textit{principal-auxiliary} bipartite structure of the
unitary, under magnitude-based parameter decoupling, one may partially transform the bare
Hamiltonian to a transformed Hamiltonian $\bar{H}_P = U_P^\dagger \hat{H} U_P$ by 
employing the Baker-Campbell-Hausdorff expansion and 
restricting the contributions of the 
\textit{auxiliary} parameters at the second power. The energy expression then takes the form -

\begin{equation} \label{Energy derivation}
\begin{split}
     E = &\bra{\Phi_{0}} \Big( \prod_{PPS}^{N_P} e^{\theta_{P}\hat{\kappa}_{P}} \prod_{APS}^{N_A}  e^{\theta_{A}(\{\theta_p\})\hat{\kappa}_{A}}\Big)^{\dagger} \hat{H} \\
     &\Big( \prod_{PPS}^{N_P} e^{\theta_{P}\hat{\kappa}_{P}} \prod_{APS}^{N_A}  e^{\theta_{A}(\{\theta_p\})\hat{\kappa}_{A}}\Big)\ket{\Phi_{0}} \\
     & =\bra{\Phi_{0}} \bar{H}_P  \ket{\Phi_{0}} + \sum_{A_\alpha \in APS} \theta_{A_\alpha} \underbrace{\bra{\Phi_{0}}  [\bar{H}_P,\hat{\kappa}_{A_\alpha}] \ket{\Phi_{0}}}_{\mbox{Term 1}}\\
    & +\frac{1}{2} \sum_{A_\alpha,A_\beta} \theta_{A_\alpha} \theta_{A_\beta} \underbrace{\bra{\Phi_{0}} \Big[[\Bar{H}_P,\hat{\kappa}_{A_\alpha}],\hat{\kappa}_{A_\beta}\Big] \ket{\Phi_{0}}}_{\mbox{Term 2}}
\end{split}
\end{equation}
Here, $\theta_A(\{\theta_P\})$ denote that the \textit{auxiliary} parameters ($\theta_A$) are not independent but rather are derived from the \textit{principal} parameters ($\theta_p$) using Eq. \eqref{thetaS_ad}. After simplifying \textit{Term 1} and \textit{Term 2} in Eq. \eqref{Energy derivation} (see Appendix \ref{app:section2}), the final energy may be written as:
% we define an 
% adiabatically decoupled unitary operator 
% \textcolor{red}{WE MAY USE THE TERM "PRINCIPAL SUB-SYSTEM UNITARY"}

% \begin{equation} \label{U_ad}
%     \begin{split}
%         \hat{U}_{ad} =& \prod_{I}^{N_P} e^{\theta_{P_I}\hat{\kappa}_{P_I}}
%     \end{split}
% \end{equation}
% \sout{With all the approximations discussed so far we can reduce the dynamical subspace for the discrete-time evolution
% but the fixed point properties, such as energy determination, still require all the parameters, resulting in a deep circuit.
% In this scenario, to make the whole protocol a reduced \textit{fixed-circuit-depth} algorithm, we can take advantage of the particular ordering
% of $\hat{U}_{pab}$ as shown in Eq. \eqref{mag based U} for an approximate energy function.
% \textcolor{blue}{Due to the principal-auxiliary bipartite structure of the unitary, under magnitude-based parameter decoupling, the BCH expansion for the auxiliary unitary 
% % in Eq. \eqref{energy with feedback}
% can be kept only to second order and
% neglecting higher order terms we can approximate (see Appendix \ref{app:section2} for the derivation) the energy equation}}
\begin{equation} \label{noisy adpqe energy equation}
    E \approx \bra{\Phi_0} \hat{U}_{P}^\dagger \hat{H} \hat{U}_{P} \ket{\Phi_0} + \sum_{\alpha=1}^{N_A} \theta_{A_{\alpha}}^2 D_{A_{\alpha}} 
\end{equation}
Eq. \eqref{noisy adpqe energy equation}, combined with Eqs. \eqref{no feedback r_PI} and \eqref{thetaS_ad} gives the entire protocol for performing \textit{no-feedback-controlled adiabatically decoupled} PQE (\textit{nfc}AD-PQE). 
% The evolution of the \textit{principal} parameters takes place according to Eq. \eqref{no feedback r_PI} followed by the post-iterative mapping using Eq. \eqref{thetaS_ad}. Once we have the entire set of converged \textit{principal} parameters and \textit{auxiliary} ones (obtained through the mapping), we can obtain the energy using Eq. \eqref{noisy adpqe energy equation}. In each of these equations, the ansatz used is $U_p$, which only involves \enquote{excitations} associated with \textit{principal} parameters.
With the formalism developed so far, the \textit{nfc}AD-PQE can be outlined with the following sequential steps-
\begin{enumerate}
    \item The initial estimate for the single and double excitation parameters are chosen 
    based on their leading order values. This implies that the doubles are estimated based on 
    their magnitudes at the first order (MP2 values) while the singles are estimated 
    from the second perturbative order\cite{halder2024noise}. Under the magnitude-based decoupling at this step, the parameter space is partitioned into \textit{PPS} and \textit{APS}. The parameter set is sorted based on their absolute magnitude while keeping track of their corresponding excitation indices. We set the value of $f_{pps}$ such that \textit{PPS} contains the largest $N_P=f_{pps}N_{par}$ parameters. The remaining $N_A = (N_{par}-N_P)$ number of parameters constitute \textit{APS}. These $N_P$ and $N_A$ number of parameters are grouped into different ranks of excitations (here singles and doubles) and are ordered to construct the unitary such that it follows the particular ordering of Eq. \eqref{mag based U}.
    \item Only the \textit{principal} parameters are updated with the residue defined in Eq. \eqref{no feedback r_PI} till convergence using
    \begin{equation} \label{iter nfcADPQE r_PI}
    \theta_{P_I}^{(k+1)} = \theta_{P_I}^{(k)} + \frac{r_{P_I}^{(k)}}{D_{P_I}}
    \end{equation}
    
    \item As the residual condition $r_{P_I} \rightarrow 0$ is satisfied, the iteration over the \textit{principal} subspace is terminated. At this fixed point in the \textit{principal}
    parameter phase space, the \textit{auxiliary} parameters are mapped via one-step \textit{post-optimization mapping} using Eq. \eqref{thetaS_ad}. Finally, the energy is
    determined via the approximated equation Eq. \eqref{noisy adpqe energy equation}. 
\end{enumerate}

In contrast to measurement-based methods like ADAPT-VQE or SPQE, \textit{nfc}AD-PQE does not discard operators from the pool but transfers them to an auxiliary subspace with a degree of approximation without needing additional selection measurements or quantum resources.
Additionally, instead of treating \textit{nfc}AD-PQE and SPQE or ADAPT-VQE 
as two competing theories, the former can be integrated with dynamic ansatz algorithms\cite{grimsley2019adaptive,stair2021simulating} for even greater resource efficiency.
% The development of \textit{nfc}AD-PQE primarily serves dual objectives. 
% First, it gives a theoretical background to produce a static shallow depth ansatz capable of capturing accurate electronic correlation.
% Second, the number of measurements required for the iterative \textit{nfc}AD-PQE is substantially lower than its conventional counterpart. Along this line, one must also 
% note that \textit{nfc}AD-PQE, being static structured, does not require any pre-circuit
% measurements towards the operator selection, and this turns out to be of immense advantage over measurement-based ansatz construction in noisy quantum devices.
% The consequential advantage of compact circuits is also evident in their 
% improved performance under noisy simulations.
In the following subsection,
we delve into a detailed discussion about the reduced resource requirements of \textit{nfc}AD-PQE and its quantifiable impact in scenarios with noise.

% we will have a detailed discussion about the reduced resource requirements
% for \textit{nfc}AD-PQE and its quantitative effect in noisy scenario.

% The development of \textit{nfc}AD-PQE serves dual objectives. First, it gives a theoretical background to produce a compact ansatz capable
% of capturing accurate electronic correlation, resulting in a shallow fixed-circuit-depth algorithm.
% Second, the number of measurements required for the iterative \textit{nfc}AD-PQE is substantially lower than its conventional counterpart. In this
% context, it must be mentioned that we
% do not discard any operator from the operator pool; rather, we transfer the seemingly \enquote{unimportant} ones in the auxiliary
% subspace and map them with a certain degree of approximation, and hence, it does not require any additional selection measurement.
% \textcolor{blue}{Additionally, \textit{nfc}AD-PQE can be integrated with dynamic ansatz algorithms\cite{grimsley2019adaptive,stair2021simulating}
% for a more efficient algorithmic implementation. In the next subsection, we will have a detailed discussion about the reduced resource requirements for \textit{nfc}AD-PQE.}

\subsection{Advantages of \textit{nfc}AD-PQE} \label{advantages nfcadpqe}
\subsubsection*{Reduction in number of measurements} \label{reduction measurements section}
The primary advantage inherent in the application of \textit{nfc}AD-PQE lies in its exclusive optimization of the \textit{principal} subset. This subspace 
optimization strategy results in a drastic reduction in the number of residue 
component evaluations, which has explicit effects on the number of measurements. Quantitatively,
the number of measurements ($m_{res}$) needed to compute the residual vector involving $N_{par}$
parameters with precision $\epsilon$ has an upper bound \cite{stair2021simulating,romero2018strategies} 

\begin{equation}
    m_{res} \leq 3N_{par} \frac{(\sum_{l} \mid h_l\mid)^2}{\epsilon^2}
\end{equation}
where $h_l$ is the coefficient of the $l$-th Pauli string in the Jordan-Wigner-mapped Hamiltonian\cite{jordan1993paulische}. Consequently, the ratio between the number of measurements utilized by \textit{nfc}AD-PQE to that of conventional PQE (per residue vector evaluation) approximately becomes 
\begin{equation}
    \frac{m_{res}^{nfcAD-PQE}}{m_{res}^{PQE}}=f_{pps} < 1
\end{equation}
which translates into 
significant reduction in measurement costs when implemented throughout the entire iterative process until convergence is achieved.
Along this line, one must also 
note that \textit{nfc}AD-PQE, being static structured, does not require any \textit{pre-circuit
measurements} towards the operator selection, and this turns out to be of immense advantage over measurement-based ansatz construction in noisy quantum devices.
% {immense curtailment in
% measurement cost when realised over the full set of iterations
% till convergence. }

% which significant reduction in measurement costs when implemented throughout the entire iterative process until convergence is achieved.

\subsubsection*{Reduction in gate depth}
Minimizing the number of CNOT gates stands as a paramount objective for any NISQ-friendly algorithm.
% The reduction of CNOT gates is one of the primary motives of any NISQ-friendly algorithm.
In the case of a serial implementation of dUCCSD ansatz
construction with Jordan-Wigner mapping, the number of gates (mainly CNOT) scales as\cite{cao2019quantum} $\sim N^5$, where $N$ is the number
of qubits.
However, with \textit{nfc}AD-PQE, where circuits are parameterized solely by the \textit{principal parameters}, the gate requirement diminishes to around $\sim f_{pps}N^5$. In practical terms, this translates to significant reductions in CNOT gates since $f_{pps}<1$.
CNOT gates are extremely susceptible to noise in hardware realizations. Therefore, reducing the number of CNOT gates ultimately enhances performance in realistic scenarios, promising more robust outcomes.

\subsubsection*{Reduced impact of noise}
The focus of the algorithm on enhanced applicability under noise motivates a discussion on the mathematical relationship between noise and circuit depth. 
%Since the algorithm targets better applicability under noise 
% and as the CNOT gates are among the most noisy \textit{fault locations} in a quantum circuit, 
% To gain further quantitative insight,
%here we discuss briefly the mathematical connection between noise and circuit depth. 
Noise can be modeled by
discrete probabilistic events called \textit{faults} that can occur at a variety of locations in a quantum circuit \cite{gottesman2010introduction,terhal2015quantum}. Considering Pauli noise
with probability $p_l$ at circuit location $l$, the probability that no fault occurs (or the \textit{fault-free probability} is \cite{cai2023quantum}) $P_0 = \prod_l (1-p_l)$.
% \begin{equation}
%     P_0 = \prod_l (1-p_l)
% \end{equation}
For a measure of the amount of noise, we can define the average
number of circuit faults per run or the \textit{circuit fault rate} $r = \sum_l p_l$.
% \begin{equation}
%     \lambda = \sum_l p_l
% \end{equation}
% It can be shown that the \textit{fault free probability} of a given circuit decays exponentially\cite{cai2023quantum} with \textit{circuit fault rate}, \textcolor{red}{i.e.}
% $P_0 = e^{-\lambda}$.
For a circuit with location-independent Pauli noise probability $p$ the circuit fault rate becomes $\lambda = Lp$, where $L$ is the total number of \textit{fault locations} and
the corresponding \textit{fault free probability} is $P_0 = e^{-Lp}$.
As we have already discussed in section \ref{noiseless_data}, the number of CNOT gates is a rough estimate of the number of \textit{fault locations}
in a circuit which scales as $\sim N^5$ for PQE and $\sim f_{pps}N^5$ for \textit{nfc}AD-PQE. The corresponding \textit{fault free probabilities} are
\begin{equation}
    \begin{split}
        & P^{(PQE)}_0 \sim e^{-N^5}\\
        & P^{(nfcAD-PQE)}_0\sim e^{-f_{pps}N^5}
    \end{split}
\end{equation}
respectively. Due to the presence of the fraction $f_{pps}$, the fault-free probability for \textit{nfc}AD-PQE $(P_0^{(nfcAD-PQE)})$ clearly decays much slower than that of PQE $(P_0^{(PQE)})$ indicating less accumulation of noise
% better \textcolor{red}{applicability} of
in \textit{nfc}AD-PQE framework for
% particularly for
NISQ platforms.

\subsubsection*{Improved efficacy towards Error Mitigation}
To obtain meaningful accuracy in computations performed in noisy quantum hardware, one must
apply additional layers of error mitigation methods. Zero Noise Extrapolation (ZNE) \cite{temme2017error,giurgica2020digital} stands out as one of the most widely used methods
to mitigate gate-based errors. It retrieves the expectation
value of an operator by extrapolating to zero 
noise limit from the data obtained through constructing the quantum circuit at various effective degrees of processor noise.  However, the efficacy of ZNE and the quality of error mitigation deteriorate
with the increasing depth of the circuit. For ZNE with $n$-th order Richardson extrapolation method, the upper
bound of the error for the zero noise estimate is $\sim \mathcal{O}((NTr)^{n+1})$ \cite{kim2023scalable,cai2021practical}, where $N$ is the number of qubits and
$T$ is the total evolution time. Therefore, for a pronounced efficacy of ZNE, a smaller $NTr$ is required. 
% As \textit{nfc}AD-PQE involves shallow depth circuits which scales as $\sim(f_{pps}N^5)$ with
% $f_{pps} < 1$, 
\textit{nfc}AD-PQE furnishes a low value of $r$ as compared to conventional PQE. Thus, for \textit{nfc}AD-PQE, the upper bound for the error in the zero noise estimate becomes $\mathcal{O}(f_{pps}^{n+1}
(NTr)^{n+1})$. Due to the conditions $f_{pps}<1$ and $n>1$, the extrapolation error bound for ZNE is analytically shown to be reduced, which translates to a more pronounced efficiency of ZNE in \textit{nfc}AD-PQE. Moreover, the \textit{sampling cost} of an error mitigation protocol is also tremendously minimized as it grows exponentially with circuit depth\cite{takagi2022fundamental}. 

Having analytically established the multi-fold advantages 
of \textit{nfc}AD-PQE implementation at a reduced quantum complexity, 
we will analyze its accuracy in the 
next section. Towards this, we shall discuss the results of quantum simulations 
under synthetic device noise as well as in ideal quantum simulators, and this will
demonstrate the superiority of the algorithm over allied methods to decipher its short
and long-term benefits.

\section{Results and Numerical Analysis}
\subsection{\textit{nfc}AD-PQE Under Noiseless Environment: Numerical Demonstrations of its Robustness for the Fault-tolerant Era} \label{noiseless_data}

\begin{figure}[!ht]
    \centering
\includegraphics[width=\linewidth]{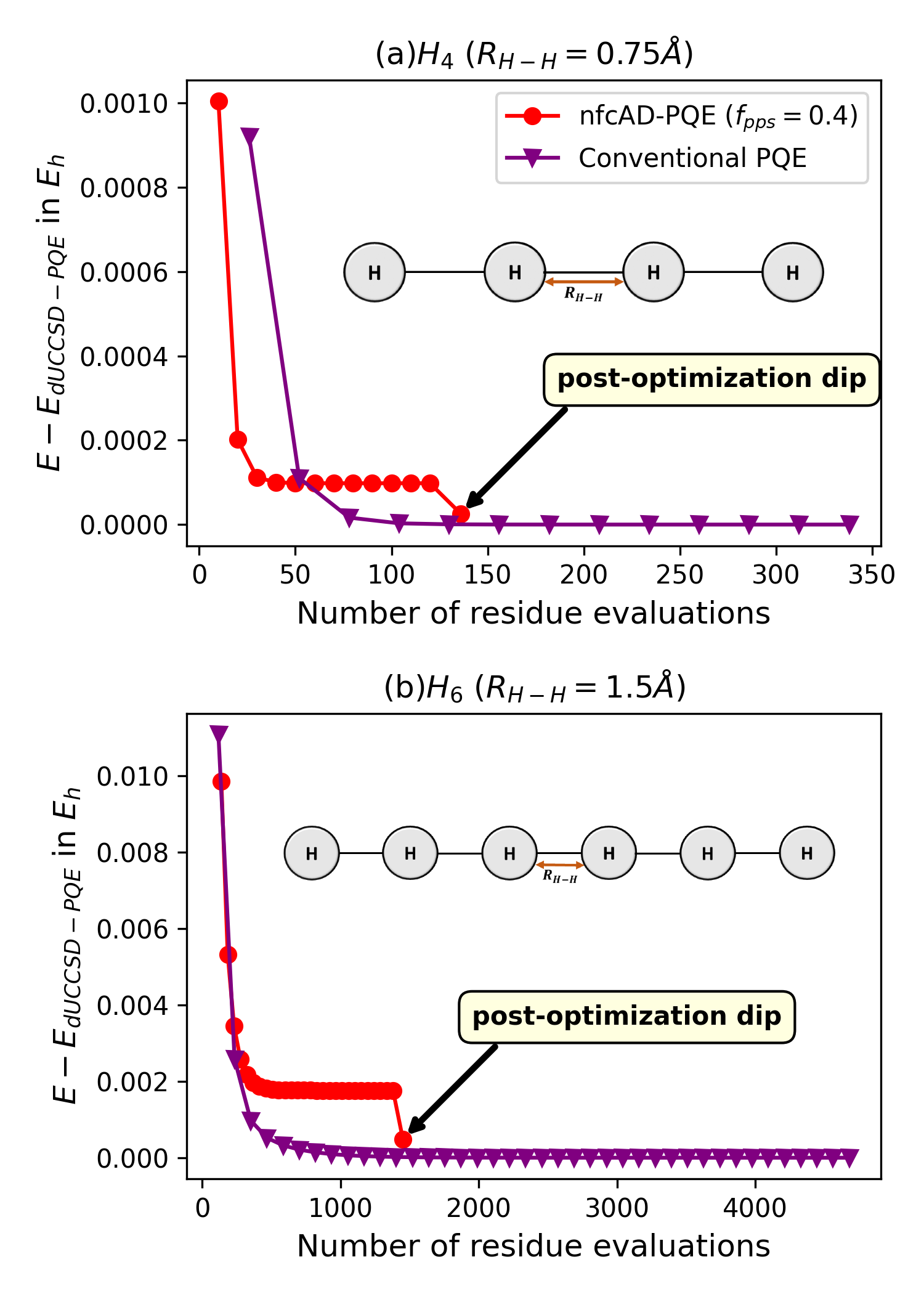}
\caption{Convergence profile of \textit{nfc}AD-PQE against the conventional 
dUCCSD-PQE  for (a) $H_4$ ($R_{H-H}=0.75$\AA) and (b) $H_6$ ($R_{H-H}=1.5$\AA) in STO-3G basis. The post-optimization dip is the signature of single-step mapping to incorporate auxiliary amplitudes.}
    \label{residue_count_accuracy_plot}
\end{figure}

\begin{figure*}[!ht]
    \centering  
\includegraphics[width=\textwidth]{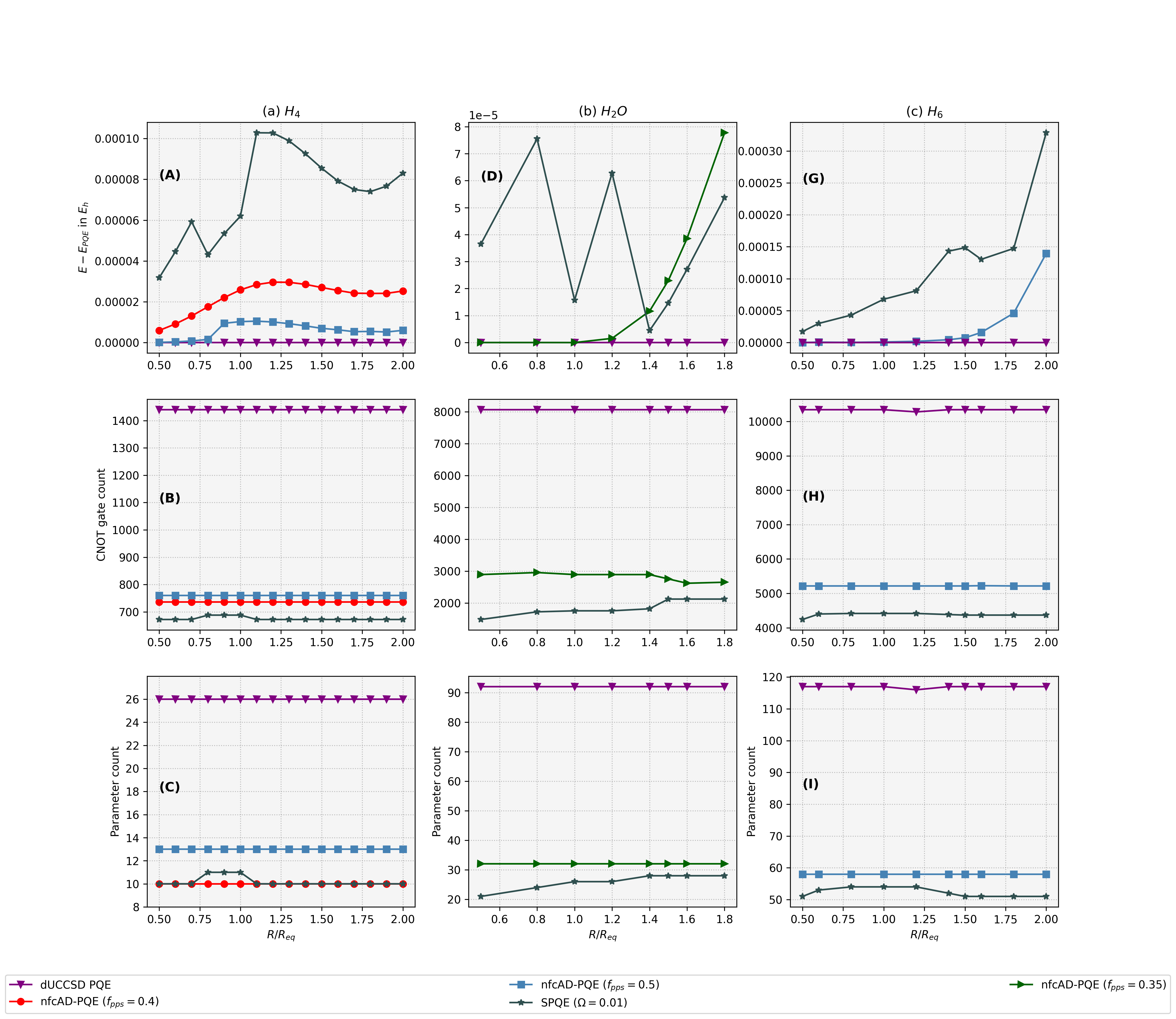}
\caption{Potential energy surface for (a)$H_4$ ($R_{(eq)H-H}=0.75$\AA), (b) $H_2O$ ($\angle H-O-H =104.4776^\circ$, $R_{(eq)O-H}=0.958$\AA) and (c)$H_6$ ($R_{(eq)H-H}=0.75$\AA) in STO-3G basis. 
The SPQE algorithm is performed with default setting (macro-iteration threshold $\Omega=0.01$ and time evolution parameter $\Delta t = 0.001$) as implemented in QForte\cite{stair2022qforte}. To keep things on an equal footing, the operator pool for SPQE is restricted to
singles and doubles. The principal fraction used for \textit{nfc}AD-PQE ranges from $f_{pps}=0.35$ to 0.5. The first row (A,D,G) represents the energy of each individual
method (\textit{nfc}AD-PQE and SPQE) with respect to dUCCSD-PQE converged energy 
($E_{PQE}$) for different inter-atomic separations.
The corresponding number of CNOT gates (B,E,H) and parameter counts (C,F,I) are also shown for each respective point along the potential energy profile.}
    \label{cx_comparison_plot}
\end{figure*}

% A comparative study between PQE and \textit{nfc}AD-PQE for dUCCSD energy convergence and accuracy in STO-3G basis for (a)$H_2O$ ($R_{H-O}=0.958$\AA), (b)$LiH$ ($R_{Li-H}=1.00$\AA), (c)linear $H_4$ ($R_{H-H}=1.125$\AA) chain. The y-axis represents the difference between energy at each step for a given algorithm and PQE converged energy and x-axis denotes the number of residue
% component evaluations of the form $\bra{\Phi_\mu}\Bar{H}\ket{\Phi_\mu}$ for arbitrary $\mu$.
% Convergence trajectory and accuracy is analyzed using $f_{pps}=0.3,0.4$ for (a),(b) and $f_{pps}=0.4,0.5$ for (c) to show the the systematic improvability in accuracy of the algorithm with increasing $f_{pps}$. For each of
% the cases the number of residue evaluations (and hence quantum measurements) is substantially lower for \textit{nfc}AD-PQE. The last step of the protocol consists of a conspicuous post-optimization dip due
% to auxiliary parameter mapping that is essential for achieving characteristic
% accuracy without any substantial resource overheads. The inset shows \textit{nfc}AD-PQE maintains a $10^{-4}$-$10^{-6}$ $E_h$ accuracy with conventional PQE across the entire potential energy surface with (a)$R_{(eq)O-H}^{(H_2O)}=0.958$\AA,(b)$R_{(eq)Li-H}^{(LiH)}=1.00$\AA,(c)$R_{(eq)H-H}^{(H_4)}=0.75$\AA.

In this section, we demonstrate the performance of \textit{nfc}AD-PQE 
under the ideal noiseless scenario that sets the ultimate limit of the 
algorithm for the fault-tolerant era.
For this, we will be comparing the accuracy of \textit{nfc}AD-PQE with respect 
to the conventional PQE which has been established as an accurate
resource-efficient alternative to VQE. We perform this study 
for a number of molecules with varied electronic complexities. All the numerical simulations are performed using in-house codes interfaced
with Qiskit\cite{Qiskit}. The molecular orbitals
and requisite integrals are imported from PySCF \cite{sun2018pyscf}. For
all our applications, we have used STO-3G basis and the rank
of the excitation operators is truncated at the 
singles and doubles (SD).
Furthermore, a convergence threshold of $10^{-5}$ for the residue norm has been used uniformly 
and no additional convergence accelerating technique (such as DIIS) is used.

To illustrate the convergence profile of 
\textit{nfc}AD-PQE, in Fig. \ref{residue_count_accuracy_plot},
we present the optimization landscape of (a) linear $H_4$ and (b) linear $H_6$ with $H-H$ distance set to be $0.75$\AA \hspace{0.5mm} and $1.5$\AA
\hspace{0.5mm} respectively
with $f_{pps}$ taken to be $0.4$. Fig. \ref{residue_count_accuracy_plot} also contains the energy trajectory for conventional PQE.
We reiterate that \textit{nfc}AD-PQE assumes the existence of a \textit{fixed point} 
within the \textit{principal} parameter subspace that closely
represents the global fixed point of the
entire set of parameters. The optimization is thus restricted within 
a disjoint principal parameter subspace, manifesting 
in a drastic reduction in the number of total residue component evaluations 
($r_\mu$) as predicted quantitatively
in section \ref{advantages nfcadpqe}.
% we present a comparison of the number of evaluations of the components of the residual vector $r_\mu$ against the difference in converged dUCCSD-PQE energy.
% % we have shown the number of evaluations of the components of the residual vector $r_\mu$ against the difference in converged dUCCSD-PQE energy. 
% We reiterate that \textit{nfc}AD-PQE assumes the existence of a fixed point within
% principal parameter subspace that closely represents the global fixed point of the
% entire set of parameters. The optimization is thus performed within a locally
% decoupled principal parameter subspace, manifesting in a drastic reduction in
% the number of total residue component evaluations as predicted quantitatively
% in section \ref{advantages nfcadpqe}. 
Furthermore, \textit{nfc}AD-PQE
crucially depends upon the \textit{post-optimization mapping} of the \textit{auxiliary} 
parameters from the \textit{principal} ones (Eq. \ref{thetaS_ad}). These two sets of
parameters collectively contribute to the energy function once the fixed 
point within the \textit{principal} subspace is reached. This manifests in
a distinctive \textit{post-optimization dip} in the energy landscape, ensuring its 
characteristic accuracy compared to conventional
dUCCSD-PQE. Moreover, it achieves this while requiring an order of magnitude fewer residue evaluations, resulting in fewer measurements
as analytically established in \ref{reduction measurements section}.
% \sout{number of residue evaluations (and thus 
% reduction in number of measurements) and the \textit{post-optimization dip} is exhibited in Fig. \ref{residue_count_accuracy_plot}.
% Such a convergence pattern is observed throughout different molecular systems at different geometries.}

To assess the accuracy, in Fig. \ref{cx_comparison_plot} we have studied the potential energy profiles of three challenging molecular systems
(a) stretching of linear $H_4$, (b) symmetric stretching of $H_2O$ and (c) 
stretching of linear $H_6$. In order to numerically validate the 
resource efficiency of \textit{nfc}AD-PQE, we have compared its accuracy
with selected PQE (SPQE) \cite{stair2021simulating} -- a measurement-based ansatz
compactification protocol within the PQE framework.  To ensure a fair comparison, we constrained the SPQE operator pool to only include single and double (SD) excitations. It was performed with an
importance selection threshold (or the macro-iteration threshold) of $\Omega=0.01$, and a time-length of $\Delta t=0.001$ for
the unitary evolution as implemented by default in QForte\cite{stair2022qforte} in conjunction with Psi4\cite{smith2020psi4}.
For a detailed discussion on SPQE, we refer to the work by Evangelista and co-workers \cite{stair2021simulating,magoulas2023cnot}.
% \sout{Since \textit{nfc}AD-PQE serves the purpose of ansatz compactification, here
% we have considered a comparison against the selected PQE\cite{stair2021simulating} (SPQE) which is a projective variant
% of ADAPT-VQE\cite{grimsley2019adaptive}-like dynamic and compact ansatz preparation protocol. To keep things numerically less-complex and on an equal footing
% we have restricted SPQE to singles and doubles excitation pool of operators with the operator selection threshold ($\Omega$=0.01)
% and the time-length ($\Delta t=0.9$) for the unitary evolution operator $e^{-i\Delta t \hat{H}}$ during the selection
% procedure.} 

The first row in Fig. \ref{cx_comparison_plot} represents the energy values of \textit{nfc}AD-PQE or SPQE with respect to
dUCCSD-PQE for different inter-atomic distances. The second and third row shows the number of CNOT gates and the number
of parameters respectively used in \textit{nfc}AD-PQE, SPQE, and conventional dUCCSD-PQE. 
% For near-equilibrium geometries of linear $H_4$, SPQE ($\Omega=0.01$) performs better than \textit{nfc}AD-PQE ($f_{pps}=0.4 \mbox{ and } 0.5$).
% While at stretched geometries, where strong correlation effects prevail, \textit{nfc}AD-PQE is more accurate as the
% choice of principal parameters along with the \textit{post-optimization mapping} recovers better electronic correlation
% even with less number of parameters and CNOT gates.
For linear $H_4$ \textit{nfc}AD-PQE ($f_{pps}=0.4$ and $f_{pps}=0.5$) and SPQE ($\Omega=0.01$) 
energy errors are within $10^{-5}-10^{-4}E_h$ with respect to dUCCSD-PQE throughout the potential energy surface (PES),
though the former takes slightly more
number of parameters and CNOT gates. This is due to the fact that in the fixed structural ordering of operators (see Eq. \eqref{mag based U})
\textit{nfc}AD-PQE contains more number of doubles in the principal subspace.
SPQE, being a dynamic ansatz compactification algorithm, utilizes slightly less number of parameters (and CNOT gates)
but at the cost of additional measurements for the selection and ordering optimization of \enquote{important} excitation operators.
In the case of $H_2O$, both \textit{nfc}AD-PQE ($f_{pps}=0.35$) and SPQE ($\Omega=0.01$) maintain $\sim 10^{-5}E_h$ accuracy with dUCCSD-PQE
for the entire PES. Similar trend is observed for $H_6$ also with \textit{nfc}AD-PQE ($f_{pps}=0.5$).
% For $H_6$ \textit{nfc}AD-PQE ($f_{pps}=0.5$) is more accurate with some more parameters and CNOT gates utilization. 
Although \textit{nfc}AD-PQE comes with a shallow
circuit with static structure and operator ordering, it is capable of capturing correlation uniformly over the potential energy profile,
as corroborated by very small
non-parallelity error (NPE). 
Here, we emphasize the fact that 
we aim to attain the characteristic accuracy of the parent ansatz 
(dUCCSD-PQE in this case) with lower quantum resources, and thus, we refrain 
from comparing against FCI. However, our approach is entirely general and 
the adaptability of the algorithm does not depend on the parameterization
of the ansatz.

\begin{figure*}[!ht]
    \centering
\includegraphics[width=\linewidth]{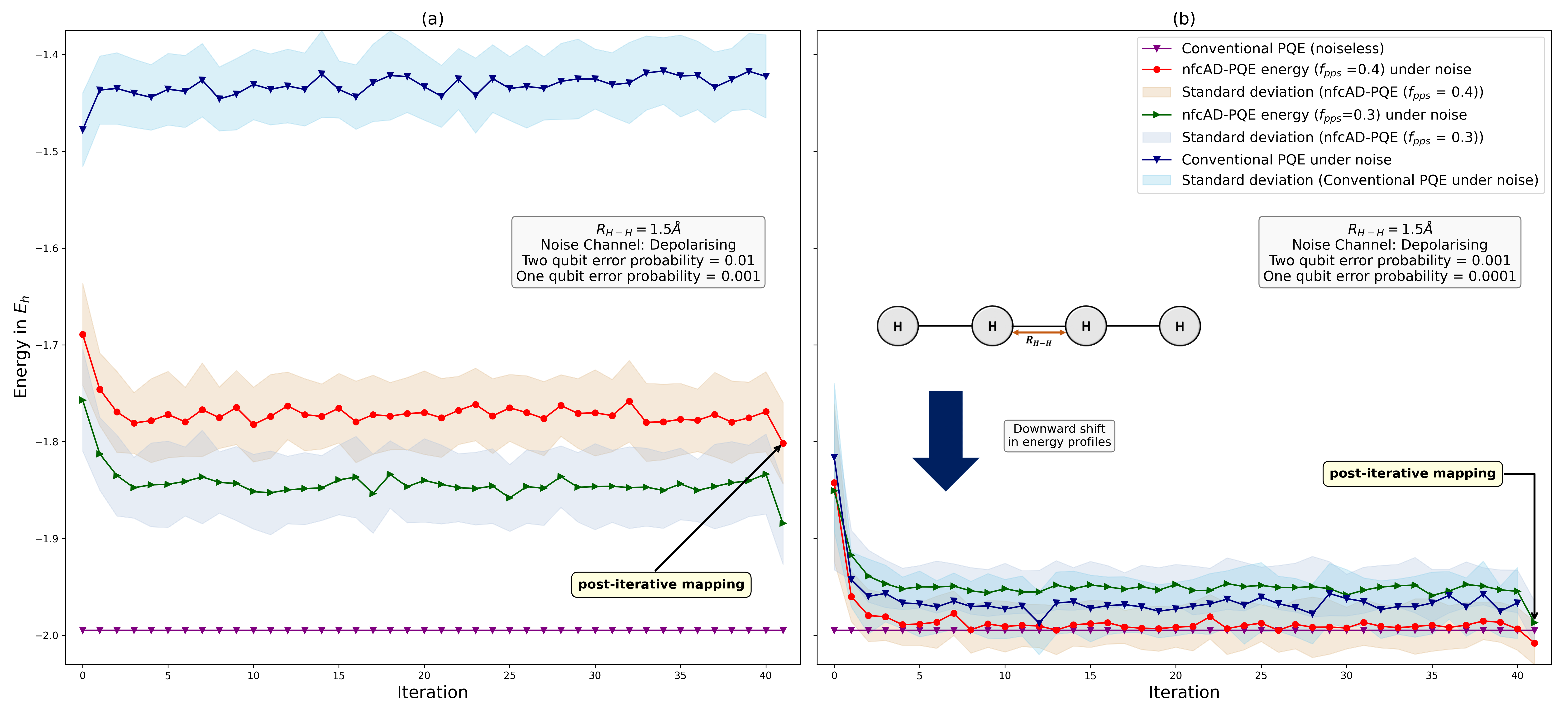}
\caption{Energy optimization landscape for \textit{nfc}AD-PQE and conventional PQE using 
dUCCSD ansatz under depolarising noise channels for linear $H_4$ ($R_{H-H}=1.5$\AA) in 
STO-3G basis. The study was performed with two different strengths of depolarising error 
parameters, as mentioned in the inset of the subplots. For every requisite expectation 
value, the number of shots was set to 5000. (a) With current hardware noise, \textit{nfc}AD-PQE shows distinct signature of convergence with significantly lower energy profiles
than dUCCSD-PQE and the characteristic 
\textit{post-optimization} dip. (b) With one order synthetic reduction of the hardware 
noise, the energy profiles for \textit{nfc}AD-PQE show qualitatively similar behavior to 
that of an ideal noiseless scenario as the electronic correlation effects start to 
predominate.
}
    \label{noisy_adpqe_image}
\end{figure*}

In summary, all of our numerical simulations in Fig. \ref{cx_comparison_plot}
demonstrate that within the range of $f_{pps}=0.3-0.5$, \textit{nfc}AD-PQE
shows an accuracy of the range $10^{-6}$-$10^{-4} E_h$ to dUCCSD-PQE, albeit 
at an order-of-magnitude reduction in the number of residue evaluations, 
number of parameters and, consequently, the CNOT gate count. In this regard, it is on par with SPQE in terms of accuracy and quantum resource utilization,
however, with 
the added advantage of zero \textit{pre-circuit measurement}-- a significant
advantage for the utility of the method in noisy devices.
% \sout{The fixed ordering of the operators (see Eq. \eqref{mag based U})
% and post-optimization mapping helps in a consistent behaviour across the entire potential energy surface.  
% This high degree of accuracy throughout the potential 
% energy profile under ideal quantum environment (that marks the fault-tolerant era)
% numerically validates the robustness 
% of the method in capturing correlation over varied electronic complexity.}
Furthermore, the accuracy of \textit{nfc}AD-PQE is
systematically improvable with $f_{pps}$, and this provides us with enough
additional flexibility to fine-tune our approach depending on the degree of 
accuracy warranted by the chemical system.
In the next section, we will numerically demonstrate 
the robustness of \textit{nfc}AD-PQE in the noisy environment.

\subsection{\textit{nfc}AD-PQE in NISQ Platforms: Numerical Simulations with Depolarizing Noise Channel}

The accuracy of \textit{nfc}AD-PQE, observed under noise-free conditions (that marks the fault-tolerant era), serves to validate 
the robustness of the underlying assumptions employed towards the genesis 
of its working equations.
% \sout{(Eq. \eqref{thetaS_ad}, \eqref{no feedback r_PI}, and \eqref{noisy adpqe energy equation})}.
However, the real advantage of this approach becomes evident when the proposed methodology is
employed in NISQ environment. At this juncture, 
one must note that the selection and the 
partitioning of the operators in the unitary ansatz are performed based on 
their magnitudes at the leading order perturbative level. Thus, unlike methods
that rely on \textit{pre-circuit measurements} to come up with 
an optimal ansatz, our ansatz is static, and it remains structurally unaffected in NISQ environment.

Fig. \ref{noisy_adpqe_image} depicts the optimization landscape of \textit{nfc}AD-PQE for 
$H_4$ ($r_{H-H}=1.5$\AA) under \textit{depolarizing noise}\cite{nielsen2010quantum} (which is one of the dominant sources 
of error in NISQ hardware). For all the computations depicted in Fig \ref{noisy_adpqe_image}, 
ZNE with random gate folding and Richardson extrapolation\cite{richardson1927viii} (with scale factors [1,2,3])
was employed throughout to compute all expectation values as an error mitigation 
protocol. It was implemented using routines from Mitiq \cite{mitiq}. While performing 
computations under noise (for both PQE and \textit{nfc}AD-PQE), we have further exploited 
qubit excitation-based (QEB) CNOT efficient quantum circuits\cite{yordanov2020efficient,magoulas2023cnot} for realizing the dUCCSD ansatz. 
This is done to further reduce the CNOT gate counts for better implementation of ZNE.
Owing to disruptive noisy fluctuations, the iterative method fails to converge to a stable fixed point.
In such a scenario, in this simulation, the iteration was terminated at
iterative counter value 40. At the termination point of
each iteration, we take the average of mitigated principal parameters for the preceding 10 iterations and with that averaged set of \textit{principal}
parameters the \textit{auxiliary} parameters are mapped. 
Given the stochastic nature of energy values at each iteration, we conducted 50 independent runs,
computing the average with the corresponding standard deviation as shown in Fig. \ref{noisy_adpqe_image}.
From Fig. \ref{noisy_adpqe_image}(a), it can be discerned how \textit{nfc}AD-PQE produces
better performance under noise than conventional dUCCSD-PQE with significantly lower 
energy profile. With present-day hardware noise,
conventional PQE shows no convergence signature, while for \textit{nfc}AD-PQE, a distinct
convergence signature is followed by a discernible \textit{post-optimization mapping} dip
(to be considered the final energy value obtained by the method).
The improved accuracy for \textit{nfc}AD-PQE may be attributed to a combined effect of 
less accumulation of noise due to the reduction in the number of \textit{fault locations} and 
lower ZNE extrapolation error bounds, as discussed in section \ref{advantages nfcadpqe}. 
Surprisingly, at the current hardware noise level, Fig. \ref{noisy_adpqe_image}(a) 
shows \textit{nfc}AD-PQE with lower $f_{pps}$ is more
accurate, contrary to the ideal environment scenario (Fig. \ref{cx_comparison_plot}).
A plausible explanation for this is that, at the current level of hardware noise, 
a shallower quantum circuit, attributed to fewer \textit{fault locations},
yields superior results even though such an ansatz is not adequately expressive to
capture accurate many-body correlation. Thus, such superior performance with a 
sub-optimally parametrized ansatz is an artifact of a high degree of hardware noise.
With an improvement in the quantum hardware, such a sub-optimally parameterized ansatz
would be less accurate, and the energy computation would be more dominated by 
electronic correlation effects, ultimately leading to the noiseless results discussed 
in section \ref{noiseless_data}. To numerically demonstrate such a transition, in 
Fig. \ref{noisy_adpqe_image}(b) we have simulated the algorithms with one order-of-magnitude 
less depolarizing error parameters.
Expectedly, the results reveal a downward shift in the overall energy profiles, 
commencing a trend that qualitatively shows the behavior illustrated in 
Fig. \ref{cx_comparison_plot}.
At this lower noise intensity the \textit{nfc}AD-PQE with $f_{pps}=0.4$, despite 
containing deeper circuits is more accurate than its other counterparts under 
consideration. Such an observation suggests that, in this case, the choice of the 
principal parameter subspace is optimal to maintain a better trade-off between 
handling errors due to noise while concurrently capturing the requisite 
electronic correlation energy.   
The one with $f_{pps}=0.3$ is still showing better average accuracy 
(after the \textit{post-optimization mapping}) compared to conventional PQE. However, 
it starts to exhibit a comparatively upward shift, indicative of its sub-optimal
\textit{principal subspace} parameterization when it comes to the energy estimation 
under ideal circumstances. Thus, it is imperative that the \textit{principal} subspace is 
optimally parameterized to strike the right balance, and our numerical investigation 
shows that $f_{pps}=0.4$ is the optimal \textit{principal fraction} in this regard. 

% limited expressiveness relative to
% even though it starts to shift upward showing its lack of expressibility compared to $f_{pps}=0.4$.

To summarize, our numerical investigations establish the imperative
that \textit{nfc}AD-PQE strikes a balance between two extremely important aspects.
It has better capability to efficiently handle noise-induced errors
for more effective integration into the currently available NISQ devices while 
being requisitely precise enough to reproduce characteristic accuracy in ideal 
noiseless conditions. With the improvement in quantum hardware in the near future, 
this balance would play a pivotal role in extracting useful information and harnessing the advantages of quantum speed-up for practical utility.

% As already established, under noise a shallower circuit shows better mitigation (and thus better energy accuracy) even though it is not adequately expressive. This is a 

%  Another interesting fact is that
% even though under noiseless scenarios, lowering \textit{PPS} percentage leads to poor correlation capture, under noisy scenarios the whole energy profile is altered,
% establishing the fact that even though a more compact ansatz performs better under noise, it is not always the case that it captures the correct correlation.
% In the context of Fig. \ref{noisy_adpqe_image} this effect can be seen as in noisy scenario \textit{PPS} $30\%$ has lower energy profile throughout the iterative
% trajectory than \textit{PPS} $40\%$ but the other way around in the ideal noiseless scenario.

% Even though none of them are even near to the chemical accuracy, in this work, as already discussed we are more interested in the relative behaviour.

\section{Conclusion and Future Outlook}

In this work, we have devised the \textit{nfc}AD-PQE formalism that has its roots in principles of \textit{synergetics} and \textit{nonlinear dynamics} for a resource-efficient projective quantum algorithm. The method considers the projective parameter optimization 
problem as discrete-time iterative dynamics and fundamentally relies upon
the \textit{adiabatic decoupling} of the dominant \textit{principal} 
and the submissive \textit{auxiliary} modes based on their hierarchical timescale
of convergence. 
The application of \textit{adiabatic elimination} enables the isolation of the \textit{principal} parameter
subspace from the remaining system components. 
This allows us to study the unitary evolution 
of the whole system in a significantly reduced dimensional subspace. We design an accurate 
theoretical technique that maps the \textit{principal} amplitudes to the \textit{auxiliary} ones with \textit{no-feedback-control}
assumption that naturally guides us to construct the energy functional with 
a shallow depth fixed structured circuit. We have analytically established that $\textit{nfc}$AD-PQE
requires fewer measurements, a reduced number of CNOT gates, and is less affected by hardware noise, which results in
better error mitigability when integrated with ZNE as compared to its conventional counterpart.
Under an ideal fault-tolerant environment, the
proposed methodology is capable of capturing balanced correlation effects for molecules with 
varied electronic complexity while concurrently maintaining requisite accuracy
in NISQ environment, albeit at a significantly lower quantum complexity and without the 
requirements of any pre-circuit quantum measurements to come up with an optimal ansatz.
% In the context of this formalism, the
% immediate advantage in this NISQ era is manifested in the application of quantum 
% hardware under noisy quantum channels. We have shown extensive studies on how better the
% \textit{no-feedback-controlled} AD-PQE behaves under noise compared to the conventional 
% theory and how quantum error mitigation technique -- ZNE can work better when integrated 
% with our formalism. However, the quest for chemical accuracy in this NISQ era is far from 
% over. 
We hope that with the development of better quantum computing infrastructures along with
efficient and more accurate quantum error mitigation protocols, \textit{nfc}AD-PQE will give 
rise to even more accurate results towards the exploration of novel chemical space.
% Till then, a lucrative future direction could be to incorporate the higher-order 
% correction in the auxiliary amplitudes for a \textit{beyond-adiabatic}
% approximation for even better accuracy and efficiency.

\section{Acknowledgments}
RM acknowledges the 
financial support from Industrial Research and
Consultancy Centre, IIT Bombay, and 
Science and Engineering Research Board, Government
of India. CP acknowledges UGC (University Grants Commission and SH thanks CSIR (Council of Scientific and Industrial Research) for their respective fellowships.

\section*{AUTHOR DECLARATIONS}
\subsection*{Conflict of Interest:}
The authors have no conflict of interest to disclose.

\section*{Data Availability}
The data is available upon reasonable request to the corresponding author.

\appendix

\section{Generalized solution of the auxiliary parameters and validity of the adiabatic approximation}
\label{validity_ad_approximation}

In this appendix, we discuss the validity of the adiabatic approximation from the most general solution for the auxiliary parameters.
The discrete-time dynamical equations for the parameters have the general form

\begin{equation} \label{del_tL_appendix}
\begin{split}
    \Delta \theta_{\xi} &= \Lambda_{\xi} \theta_{\xi}+ M_{\xi}(\{\theta_{P},\theta_{A}\}) %; \forall I= 1, 2, ..., N_P
\end{split}
\end{equation}
Here, diagonal and off-diagonal terms can be expressed by:
\begin{equation} \label{lambda def _appendix}
\begin{split}
    & \Lambda_{\xi} = \frac{1}{D_{\xi}} \bra{\Phi_{\xi}}  [\hat{H},\hat{\kappa}_{\xi}] \ket{\Phi_{0}}\\
    %& \Lambda_{P_I} = \frac{1}{D_{P_I}} \bra{\Phi_{P_I}}  [\hat{H},\hat{\kappa}_{P_I}] \ket{\Phi_{0}}
\end{split}
\end{equation}
% and $M_{P_I}(\{\theta_{P},\theta_{A}\})$, $M_{A_\alpha}(\{\theta_P,\theta_A\})$ contain all the nonlinear,
% off-diagonal and coupling terms:
\begin{equation} \label{P_appendix}
    \begin{split}
        &M_{\xi}(\{\theta_{P},\theta_{A}\}) = \frac{1}{D_{\xi}} \bra{\Phi_{\xi}} \hat{H} + \sum_{\nu \neq \xi} \theta_{\nu} [\hat{H},\hat{\kappa}_{\nu}] \\
 & + \sum_{\nu} \sum_{\mu > \nu} \theta_{\nu}
 \theta_{\mu} \Big[[\hat{H},\hat{\kappa}_{\nu}],\hat{\kappa}_{\mu}\Big] + . . . \ket{\Phi_0}
    \end{split}
\end{equation}
where, depending on the context $\xi \in$ \textit{PPS} or \textit{APS}. Note that in both the equations above, we have not yet imposed any restrictions on the labels $\mu,\nu$. The most general solution to the \textit{auxiliary} parameter equation Eq. \eqref{del_thetaS} can be immediately written as \cite{haken1997discrete}

\begin{equation}\label{theta_Ai gen_appendix}
    \theta_{A_\alpha} = \sum_{m=-\infty}^{l} (1+ \Lambda_{A_\alpha})^{l-m} M_{A_\alpha}
\end{equation}
By definition $\Lambda_{A_\alpha}$ (Eq. \eqref{lambda def _appendix}) is in general negative (see Appendix \ref{app:section1}), ensuring the convergent series expansion of the \textit{auxiliary} modes.
Starting from Eq. \eqref{theta_Ai gen_appendix}, we can employ \textit{summation by parts} to further simplify\cite{haken1997discrete}-
\begin{equation} \label{expanded_theta_A_appendix} 
\begin{split}
  & \theta_{A_\alpha} = M_{A_\alpha} \sum_{m=-\infty}^{l} (1+ \Lambda_{A_\alpha})^{l-m} \\ 
  & - \sum_{m=-\infty}^{l} (1+ \Lambda_{A_\alpha})^{l+1-m} \sum_{m'=-\infty}^{m-1} (1+ \Lambda_{A_\alpha})^{m-1-m'} \Delta M_{A_\alpha} 
\end{split}
\end{equation}
Carrying out the summations in Eq. \eqref{expanded_theta_A_appendix} leads to the final solution for the auxiliary modes %neglecting the relatively smaller higher order terms
\begin{equation} \label{ts_gen_full_appendix}
    \theta_{A_\alpha} = -\frac{M(\{\theta_P,\theta_A\})}{\Lambda_{A_\alpha}} -\frac{\Delta M(\{\theta_P,\theta_A\})}{\Lambda_{A_\alpha}^2}    
\end{equation}
However, due to the nature of convergence, the time variation of the auxiliary modes can be neglected
(i.e. $\Delta{\theta_{A_\alpha}}=0$ in Eq. \eqref{del_thetaS})
via \textit{adiabatic approximation}\cite{haken1983nonlinear,agarawal2021approximate,patra2023synergistic,van1985elimination} in the characteristic timescale of
convergence of the \textit{principal} parameters.
Starting from Eq. \eqref{del_thetaS} we can employ \textit{adiabatic approximation} to immediately get the \textit{auxiliary} mode solution 
\begin{equation} \label{ts_gen_appendix}
    \theta_{A_\alpha} = -\frac{M(\{\theta_P,\theta_A\})}{\Lambda_{A_\alpha}} %-\frac{\Delta M(\{\theta_P,\theta_A\})}{\Lambda_{A_\alpha}^2}    
\end{equation}
Further, the relative magnitude condition (Eq. \eqref{magnitude params}) can be invoked to express the \textit{auxiliary} parameters as a function of \textit{principal} parameters only
\begin{equation} \label{thetaS_ad_appendix_1}
    \begin{split}
        \theta_{A_\alpha} &= -\frac{M(\{\theta_P,\theta_A\})}{\Lambda_{A_\alpha}} \xrightarrow{{\mid \theta_A \mid } << {\mid \theta_P \mid }} {-\frac{M_{A_\alpha}(\{\theta_P\})}{\Lambda_{A_\alpha}}}\\
    \end{split}
\end{equation}
From Eq. \eqref{P_appendix} one can write $M_{A_\alpha}(\{\theta_P\})$ explicitly as 
\begin{equation} \label{M_as_func_of_thetaP}
    \begin{split}
        &M_{A_\alpha}(\{\theta_{P}\}) \approx \frac{1}{D_{A_\alpha}} \bra{\Phi_{A_\alpha}} \hat{H} + \sum_{P_I} \theta_{P_I} [\hat{H},\hat{\kappa}_{P_I}] \\
 & + \sum_{P_I} \sum_{P_J > P_I} \theta_{P_I}
 \theta_{P_J} \Big[[\hat{H},\hat{\kappa}_{P_I}],\hat{\kappa}_{P_J}\Big] + . . . \ket{\Phi_0}\\
        & = -\frac{\bra{\Phi_{A_\alpha}} \hat{U}_{P}^{\dagger}(\{\theta_P\}) \hat{H} \hat{U}_{P}(\{\theta_P\}) \ket{\Phi_0}}{\bra{\Phi_{A_\alpha}}  [\hat{H},\hat{\kappa}_{A_\alpha}] \ket{\Phi_{0}}}
    \end{split}
\end{equation}
Using this Eq. \eqref{thetaS_ad_appendix_1} can be compactly written as-
\begin{equation} \label{thetaS_ad_appendix}
    \begin{split}
        \theta_{A_\alpha} &= -\frac{\bra{\Phi_{A_\alpha}} \hat{U}_{P}^{\dagger}(\{\theta_P\}) \hat{H} \hat{U}_{P}(\{\theta_P\}) \ket{\Phi_0}}{\bra{\Phi_{A_\alpha}}  [\hat{H},\hat{\kappa}_{A_\alpha}] \ket{\Phi_{0}}} \\
        & = \frac{\bra{\Phi_{A_\alpha}} \hat{U}_{P}^{\dagger}(\{\theta_P\}) \hat{H} \hat{U}_{P}(\{\theta_P\}) \ket{\Phi_0}}{D_{A_\alpha}}
    \end{split}
\end{equation}
where, $\hat{U}_P(\{\theta_P\})$ is defined in Eq. \eqref{U_P} and ${\bra{\Phi_{A_\alpha}}  [\hat{H},\hat{\kappa}_{A_\alpha}] \ket{\Phi_{0}}} \approx -D_{A_\alpha}$ under leading order approximation
(see Appendix \eqref{app:section1} for the validity of the approximation).
A direct comparison of Eq. \eqref{thetaS_ad_appendix}
with \eqref{ts_gen_full_appendix} reveals that the \textit{adiabatic approximation} (Eq. \eqref{thetaS_ad_appendix}) constitutes the leading order term of the generalized solution,
indicating its considerable accuracy \cite{wunderlin1981generalized}.

\section{Approximation of the commutator terms and convergence criteria for series expansion of auxiliary mode solutions }
\label{app:section1}

In this section, we provide a mathematical rationale behind the convergence of the most general solution (Eq. \eqref{theta_Ai gen_appendix}) along with the validity
of the particular compact structure of the denominator in Eq. \eqref{thetaS_ad}. 
The commutator of the form $\bra{\Phi_{\mu}}  [\hat{H},\hat{\kappa}_{\nu}] \ket{\Phi_{0}}$ can be
approximated by expanding the Hamiltonian into a zeroth order one-body Fock operator and first order two-body operator $\hat{H}=\hat{F}^{(0)} + \hat{V}^{(1)}$
and keeping the zeroth order term only
\begin{equation} \label{commutator apprximation appendix}
\begin{split}
    & \bra{\Phi_{\mu}}  [\hat{H},\hat{\kappa}_{\nu}] \ket{\Phi_{0}}  \\
    \implies & \bra{\Phi_{\mu}}  [\hat{F}^{(0)},\hat{\kappa}_{\nu}] \ket{\Phi_{0}} \approx -D_{\mu} \delta_{\mu \nu}
\end{split}
\end{equation}

The definition of the coefficient of the linear terms has the generic form-
\begin{equation} \label{lambda general def}
\begin{split}
    & \Lambda_{\mu} = \frac{1}{D_{\mu}} \bra{\Phi_{\mu}}  [\hat{H},\hat{\kappa}_{\mu}] \ket{\Phi_{0}}
\end{split}
\end{equation}

From the results of Eq. \eqref{commutator apprximation appendix} we can see that $\Lambda_{\mu}$ is negative and $\mid \Lambda_\mu \mid \sim 1$ such that $\mid (1+\Lambda_{\mu}) \mid << 1$ ensuring the
convergence of the series in Eq. \eqref{theta_Ai gen_appendix}.

\section{Derivation of the low depth energy determining equation}
\label{app:section2}
Here, we analytically establish the energy expression (Eq. \eqref{noisy adpqe energy equation}) for a shallow-depth utilization.
With the \textit{principal-auxiliary bipartite} operator (Eq. \eqref{mag based U}), the energy term can be expressed by 
\begin{equation} \label{energy with feedback}
   E_{AD-PQE}(\boldsymbol{\theta}) = \bra{\Phi_o}\hat{U}^{\dagger}_{pab}(\boldsymbol{\theta})H \hat{U}_{pab}(\boldsymbol{\theta})\ket{\Phi_o}
\end{equation}

% that takes advantage of the bipartite form
% of the operator $\hat{U}_{pab}$ - 

% \begin{equation} \label{Energy derivation}
% \begin{split}
%     & E = \bra{\Phi_{0}} \Big( \prod_{PPS}^{N_P} e^{\theta_{P}\hat{\kappa}_{P}} \prod_{APS}^{N_A}  e^{\theta_{A}\hat{\kappa}_{A}}\Big)^{\dagger} \hat{H} \Big( \prod_{PPS}^{N_P} e^{\theta_{P}\hat{\kappa}_{P}} \prod_{APS}^{N_A}  e^{\theta_{A}\hat{\kappa}_{A}}\Big)\ket{\Phi_{0}} \\
%     & \implies = \bra{\Phi_{0}} \bar{H}_P  \ket{\Phi_{0}} + \sum_{A_\alpha \in APS} \theta_{A_\alpha} \underbrace{\bra{\Phi_{0}}  [\bar{H}_P,\hat{\kappa}_{A_\alpha}] \ket{\Phi_{0}}}_{\mbox{Term 1}} \\
%     & +\frac{1}{2} \sum_{A_\alpha,A_\beta} \theta_{A_\alpha} \theta_{A_\beta} \underbrace{\bra{\Phi_{0}} \Big[[\Bar{H}_P,\hat{\kappa}_{A_\alpha}],\hat{\kappa}_{A_\beta}\Big] \ket{\Phi_{0}}}_{\mbox{Term 2}} +.... \mbox{h.o.t.}
% \end{split}
% \end{equation}
Term 1 and Term 2 in Eq. \eqref{Energy derivation} can be further expanded-

\subsubsection*{Term 1}

\begin{equation} \label{Energy derivation Term 1}
\begin{split}
    & \bra{\Phi_{0}}  [\bar{H}_P,\hat{\kappa}_{A_\alpha}] \ket{\Phi_{0}} \\
    & = \bra{\Phi_{0}} \Bar{H}_P \hat{\kappa}_{A_\alpha} \ket{\Phi_0} - \bra{\Phi_{0}} \hat{\kappa}_{A_\alpha} \Bar{H}_P  \ket{\Phi_0} \\
    & = \bra{\Phi_{0}} \Bar{H}_P \ket{\Phi_{A_\alpha}} + \bra{\Phi_{A_\alpha}}  \Bar{H}_P  \ket{\Phi_0} \\
    & = 2 \bra{\Phi_{A_\alpha}} \hat{U}_P^{\dagger} \hat{H} \hat{U}_P \ket{\Phi_0} \\
    & = 2 \theta_{A_\alpha} D_{A_\alpha}
\end{split}
\end{equation}
where, at the last step we have used Eq. \eqref{thetaS_ad} to replace the operator expectation value with adiabatically obtained auxiliary parameters.
\subsubsection*{Term 2}
Since term 2 is a nonlinear term, for its evaluation we first approximate the BCH expansion for $\Bar{H}_P$ up to zeroth order and then follow the
second quantized operator algebra for the final compact form-
\begin{equation} \label{Energy derivation Term 2}
\begin{split}
    & \bra{\Phi_{0}} \Big[[\Bar{H}_P,\hat{\kappa}_{A_\alpha}],\hat{\kappa}_{A_\beta}\Big] \ket{\Phi_{0}} \\
    & \approx \bra{\Phi_{0}} \Big[[\hat{H},\hat{\kappa}_{A_\alpha}],\hat{\kappa}_{A_\beta}\Big] \ket{\Phi_{0}}\\
    & = \bra{\Phi_{0}}[\hat{H},\hat{\kappa}_{A_\alpha}] \hat{\kappa}_{A_\beta} \ket{\Phi_{0}} - \bra{\Phi_{0}} \hat{\kappa}_{A_\beta} [ \hat{H},\hat{\kappa}_{A_\alpha}] \ket{\Phi_{0}}\\
    & = \bra{\Phi_{0}}[\hat{H},\hat{\kappa}_{A_\alpha}] \ket{\Phi_{A_\beta}} + \bra{\Phi_{A_\beta}} [ \hat{H},\hat{\kappa}_{A_\alpha}] \ket{\Phi_{0}} \\
    & = 2 \bra{\Phi_{A_\beta}} [ \hat{H},\hat{\kappa}_{A_\alpha}] \ket{\Phi_{0}} \\
    & = -2 D_{A_\beta} \delta_{A_\alpha A_\beta}
\end{split}
\end{equation}

Plugging the final forms of Term 1 \eqref{Energy derivation Term 1} and Term 2 \eqref{Energy derivation Term 2}  in Eq. \eqref{Energy derivation} we get

\begin{equation} \label{final energy without feedback}
\begin{split}
    & E_{AD-PQE} = \bra{\Phi_{0}} \bar{H}_P  \ket{\Phi_{0}} + \sum_{A_\alpha} 2 \theta_{A_\alpha}^2 D_{A_\alpha} \\
    & -\frac{1}{2} \sum_{A_\alpha,A_\beta} 2\theta_{A_\alpha} \theta_{A_\beta} D_{A_\beta} \delta_{A_\alpha A_\beta} \\
    & = \bra{\Phi_{0}} \bar{H}_P  \ket{\Phi_{0}} + \sum_{A_\alpha} 2 \theta_{A_\alpha}^2 D_{A_\alpha} - \sum_{A_\alpha} \theta_{A_\alpha}^2 D_{A_\alpha} \\
    & \implies E_{AD-PQE} = \bra{\Phi_{0}} \hat{U}_P^{\dagger} \hat{H} \hat{U}_P  \ket{\Phi_{0}} + \sum_{A_\alpha} \theta_{A_\alpha}^2 D_{A_\alpha}
\end{split}
\end{equation}
which leads to Eq. \eqref{noisy adpqe energy equation} that we use for the energy evaluation during the post-optimization mapping.

\section*{References:}

%\bibliography{./literature}

%aipnum4-2.bst 2019-01-14 (MD) hand-edited version of apsrev4-1.bst

% \bibliography{./literature}

%aipnum4-2.bst 2019-01-14 (MD) hand-edited version of apsrev4-1.bst

%\section{APPENDIX:}
%
%\begin{multline}
%    H_{S_i}^d = (1 + P(i,j)P(a,b)) \Big(-f_{ii}+f_{aa}+\frac{1}{2}v_{ab}^{ab}+2v_{ia}^{ai}\\-(1+\delta_{ij}\delta_{ab}) v_{ia}^{ia}+ \frac{1}{2}v_{ij}^{ij}- v_{ib}^{ib} \Big)\\
%    S_i = \{ijab\} \in \textcolor{red}{SAS}
%\end{multline}    

%\bibliography{./literature}
    
\end{document}